\documentclass[cernpreprint,txfonts,UKenglish,texlive=2011,subfigure=true,biblatex=false,texmf]{atlasdoc}
\pdfoutput=1

\usepackage{atlaspackage} 
\usepackage{atlasphysics}
\usepackage{multirow}
\usepackage{subfigure}
\usepackage[numbers,sort&compress]{natbib}

\newcommand*{\myalign}[2]{\multicolumn{1}{#1}{#2}}

\newcommand{\roots}{\sqrt{s}}

\newcommand{\qq}{{\ensuremath{q\bar{q}}}}
\newcommand{\VR}{\ensuremath{V_{\text{R}}}}
\newcommand{\WR}{\ensuremath{W_{\text{R}}}}
\newcommand{\ZR}{\ensuremath{Z'}}
\newcommand{\mWR}{\ensuremath{m_{W_{\text{R}}}}}
\newcommand{\mVR}{\ensuremath{m_{V_{\text{R}}}}}
\newcommand{\mZR}{\ensuremath{m_{Z'}}}
\newcommand{\mN}{\ensuremath{m_{N}}}

\newcommand{\lpm}{\ell^\pm}

\newcommand{\rarr}{\rightarrow}

\newcommand{\ackname}{ \noindent {\large \bf Acknowledgements}   $ $\\}

\hypersetup{pdftitle={Search for heavy Majorana neutrinos 
with the ATLAS \\detector in $pp$ collisions at $\sqrt{s} = 8 \TeV$},pdfauthor={The ATLAS Collaboration}}

\AtlasTitle{Search for heavy Majorana neutrinos 
with the ATLAS \\detector in $pp$ collisions at $\sqrt{s} = 8 \TeV$}

\author{The ATLAS Collaboration}

\AtlasRefCode{EXOT-2012-24}
\AtlasJournal{JHEP}
\PreprintIdNumber{CERN-PH-EP-2015-070}

\AtlasAbstract{%
  A search for heavy Majorana neutrinos in events containing a pair of
high-$p_{\mathrm{T}}$ leptons of the same charge and high-$p_{\mathrm{T}}$ jets is presented. The search uses 20.3~$\ifb$ of $pp$ collision data 
collected with the ATLAS detector at the CERN Large Hadron Collider
with a centre-of-mass energy of $\roots = 8 \TeV$. 
The data are found to be consistent with the background-only hypothesis based on the Standard Model expectation.
In the context of a Type-I seesaw mechanism, limits are set on the production cross-section times branching ratio for production of heavy Majorana neutrinos in the mass range between 100 and 500~GeV. 
The limits are subsequently interpreted as limits on the mixing between the heavy Majorana neutrinos and the Standard Model neutrinos. 
In the context of a left-right symmetric model, limits on the production cross-section times branching ratio are set with respect to the masses of heavy Majorana neutrinos and heavy gauge bosons $W_{\mathrm{R}}$~and $Z'$.}

\begin{document}

\maketitle

\tableofcontents

\section{Introduction}

The discovery of mixing between generations of neutrinos~\cite{Alberico:2003kd} has established that at least two of the neutrinos have small non-zero masses.
A unique feature of neutrinos compared to other fermions in the Standard Model (SM) is that neutrinos could be their own anti-particles, so-called Majorana fermions. If this is realised
in nature, then the unusually low mass scale of the light neutrinos could be generated by a seesaw mechanism~\cite{Minkowski1977421,VanNieuwenhuizen:1979hm,Mohapatra:1979ia,Mohapatra:1980yp,Davidson:1987tr,Schechter:1980gr}  
which would imply the existence of yet unobserved heavy Majorana
neutrino states. The nature of Majorana neutrinos would explicitly
allow for lepton number violation. 

In this paper, a search is presented for heavy Majorana neutrinos using the ATLAS detector at the Large Hadron Collider (LHC).
The data sample was collected in 2012 during $\roots = 8 \TeV$~$pp$~collisions and corresponds to an
integrated luminosity of $20.3\ \ifb$. Heavy Majorana neutrinos with masses above $50\GeV$~are considered. In this regime,
the production and subsequent decay of heavy Majorana neutrinos
could lead to a final state containing exactly two charged leptons, where the leptons may have the same or opposite charge in equal fractions of the heavy neutrino decays.  Only lepton pairs of the same charge (same-sign) are 
considered as there is a smaller expected SM background for pairs of same-sign leptons than for pairs of~leptons~of opposite charge (opposite-sign).
The search includes the $ee$ and $\mu\mu$~final states.

\begin{figure}[htbp]
\centering
    \includegraphics[width=0.65\textwidth]{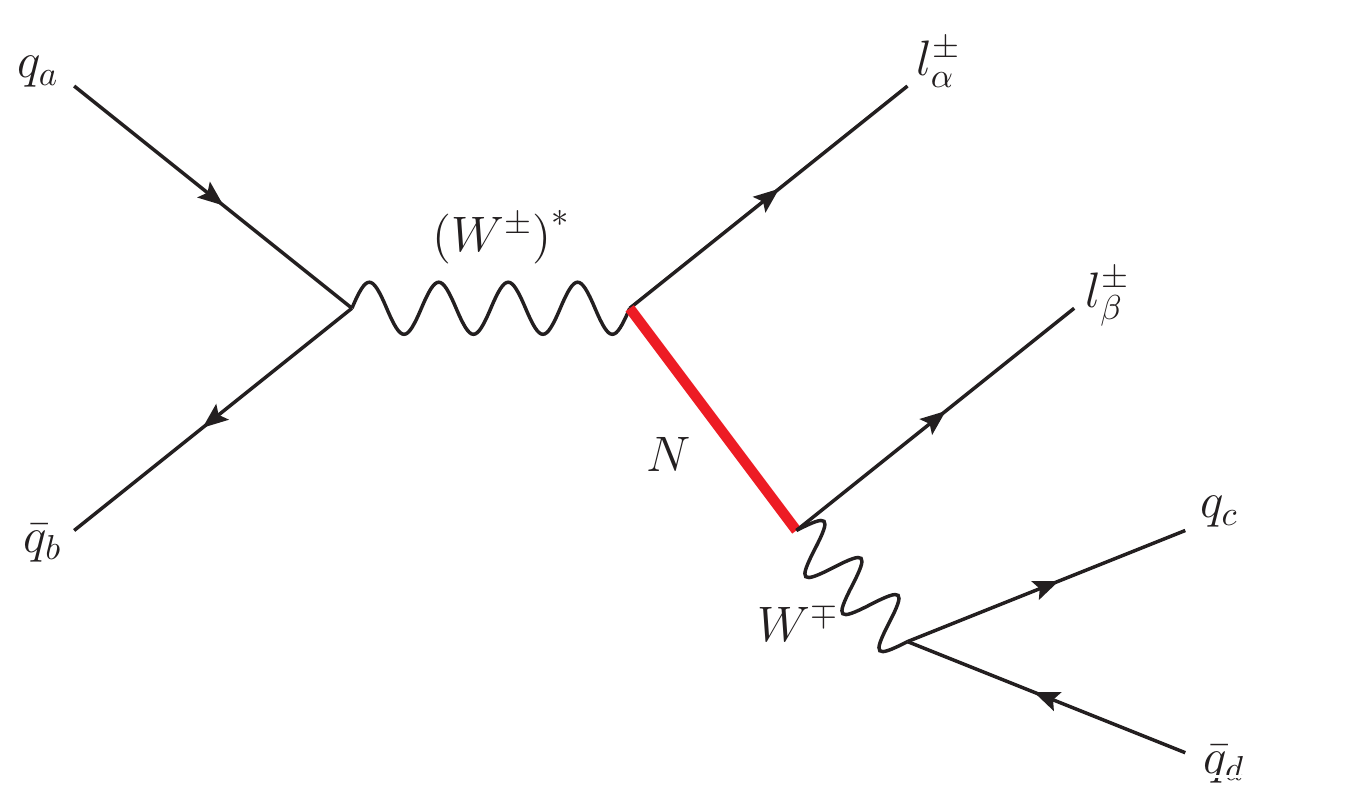}
    \caption{The tree-level diagram for the production of a heavy Majorana neutrino ($N$) in the mTISM model.
    Lepton flavour is denoted by $\alpha$ and $\beta$. Lepton flavour is assumed to be conserved, such that $\alpha = \beta$. The $W$
    boson produced from the $N$ decay is on-shell and, in this case, decays hadronically.}
    \label{fig:Feyn_Wstar}
\end{figure}

The search is guided by two theoretical models.  In the
first model,
the SM is extended in the simplest way to include right-handed neutrinos~\cite{Atre:2009rg},
such that light neutrino masses are generated by a Type-I seesaw mechanism or by radiative corrections~\cite{Pilaftsis:1991ug}.
In this minimal Type-I seesaw mechanism (mTISM),
the heavy Majorana neutrinos, $N$, can be produced via an off-shell $W$ boson, $pp \to (W^{\pm})^* \to \ell^{\pm}N$. Due to previous limits~\cite{Abreu:1996pa,Adriani:1992pq}, the 
heavy neutrino is assumed to be more massive than the $W$ boson and therefore
subsequently decays to an on-shell $W$ boson and a lepton. The on-shell $W$ boson produced in the decay of the heavy neutrino predominantly decays into 
a quark--antiquark ($\qq$) pair.
The final state in this case contains two opposite-sign or same-sign leptons
and at least two high-$\pt$ jets, where $\pt$ is the transverse momentum with respect to the beam direction.\footnote{
ATLAS uses a right-handed coordinate system, with its origin at the nominal interaction point 
in the centre of the detector. The $z$-axis points along the beam direction, the $x$-axis from 
the interaction point to the centre of the LHC ring, and the $y$-axis upwards. In the transverse plane, cylindrical 
coordinates $\left(r,\phi\right)$ are used, where $\phi$ is the azimuthal angle around the beam direction. The 
pseudorapidity $\eta$ is defined via the polar angle $\theta$ as $\eta = - \ln \tan \left(\theta/2\right)$
.} 
The tree-level process is illustrated in figure~\ref{fig:Feyn_Wstar}.
The free parameters in this model are the mixing between the heavy Majorana neutrinos and the Standard Model neutrinos, $V_{\ell N}$, and the masses of the heavy neutrinos, $m_N$.
In this framework, LEP has set direct limits for $m_{N}<m_Z$~\cite{Abreu:1996pa,Adriani:1992pq} and
CMS has set direct limits for $90 < m_N < 200 \GeV$ in $ee$ final states~\cite{Chatrchyan:2012fla} 
and $40 < m_N < 500 \GeV$ in $\mu\mu$ final states~\cite{Khachatryan:1982650}.

The second model is the left-right symmetric model (LRSM)~\cite{Pati:1974yy,theory_lrs_1,theory_lrs_2,Mohapatra:1979ia}, where a right-handed symmetry SU(2)$_{\text{R}}$ 
is added to the SM. The symmetry SU(2)$_{\text{R}}$~is assumed to be the right-handed analogue of the SM SU(2)$_{\text{L}}$~symmetry. In this model, heavy gauge bosons $\VR = \left\{ \WR , \ZR\right\}$
are also predicted and, in this analysis, the heavy gauge bosons are assumed to be more massive than the heavy neutrinos, such that they are kinematically allowed to decay into final states including heavy neutrinos.
These can be produced in the decays of heavy gauge bosons according to $\WR \to N \ell$~and $\ZR \to N N$ and
can subsequently decay via an off-shell $\WR$ boson into a lepton and a $\qq$ pair, $N \to \ell \WR^*$~with $\WR^* \to q\bar{q'}$.
The tree-level processes are shown in figure \ref{fig:Feyn_LRSM}. A previous ATLAS search in this framework has excluded $\mWR < 2.3 \TeV$ for $\mWR
 - \mN > 0.3 \TeV$ at 95\% confidence level (CL)~\cite{ATLAS_LRSM}. A more recent search performed by CMS has excluded $\mWR < 3.0$ \TeV for $\mWR
 - \mN > 0.05 \TeV$ at 95\% CL~\cite{cms_lrsm}. There are no such limits
 for the production of heavy neutrinos from $\ZR$ boson decays.

\begin{figure}[t]
\centering
    \includegraphics[width=0.49\textwidth]{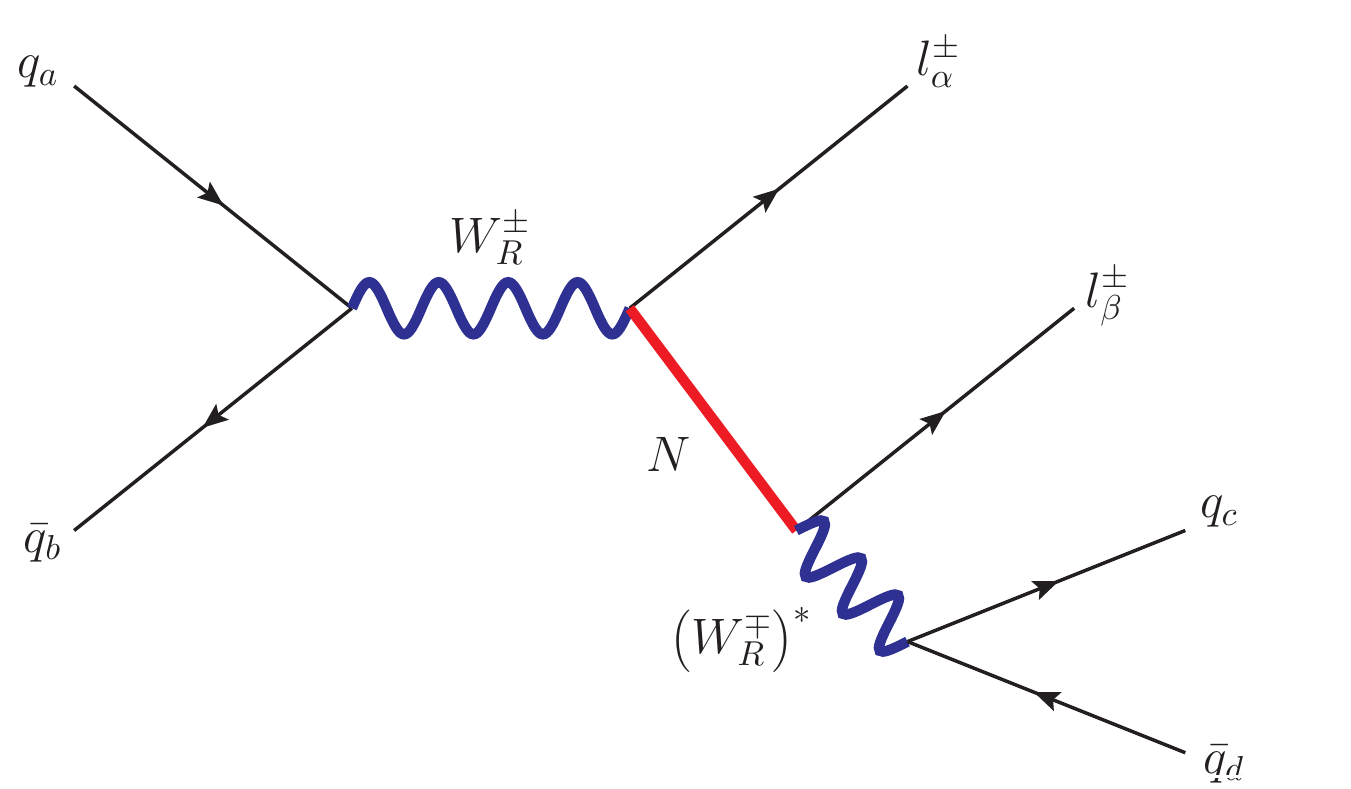}
    \includegraphics[width=0.49\textwidth]{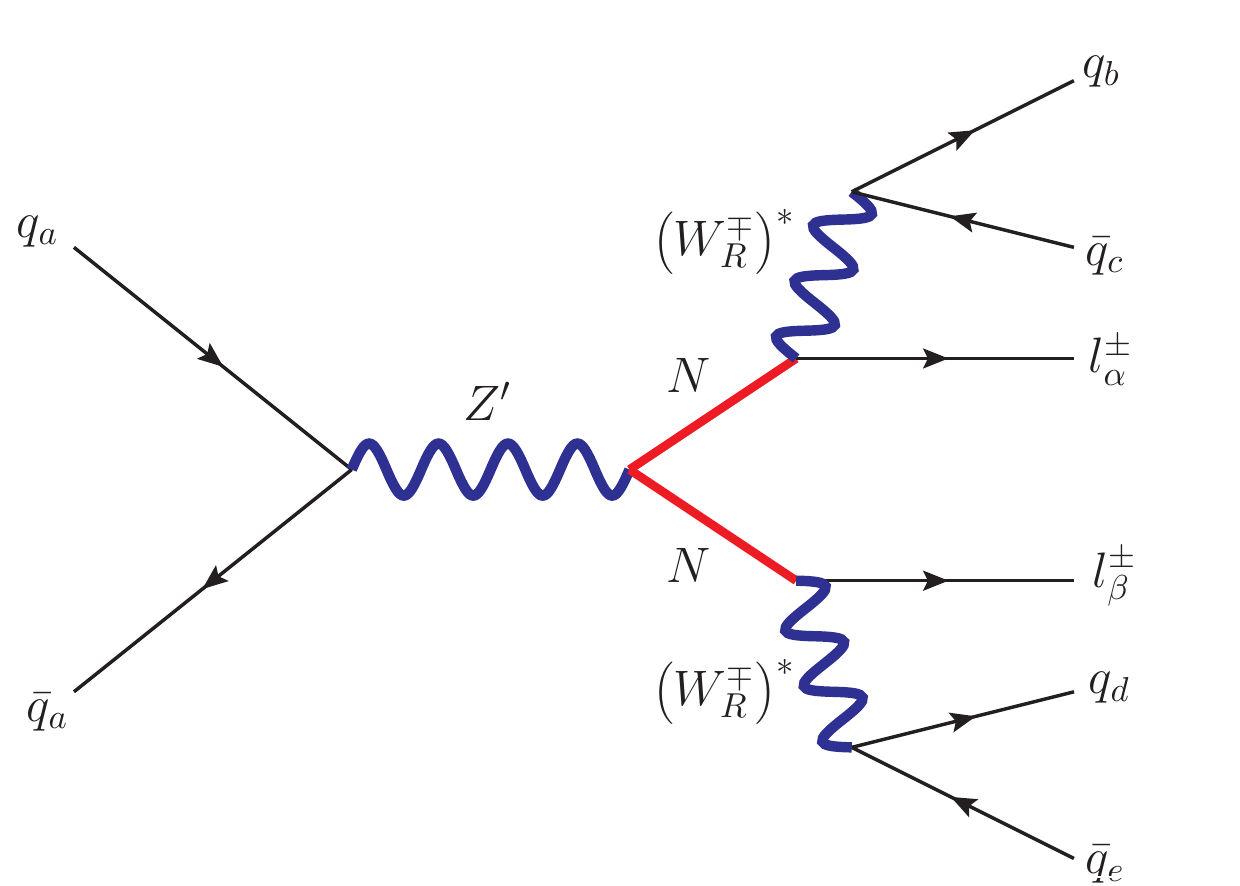}
    \caption{The tree-level diagrams for the production of a heavy Majorana neutrino ($N$) in the LRSM model, in which heavy gauge bosons $W_{\text{R}}$ and $\ZR$
    are also incorporated. Lepton flavour is denoted by $\alpha$ and $\beta$.
    Lepton flavour is assumed to be conserved, such that $\alpha = \beta$. The $W_{\text{R}}$ boson produced from the $N$ decay is off-shell and, in this case, decays hadronically.}
    \label{fig:Feyn_LRSM}
\end{figure}

Both the mTISM and LRSM models produce final states containing two same-sign leptons and high-$\pt$ jets, 
but the kinematic characteristics of the events are quite different. In the mTISM final state, one can reconstruct the resonant SM $W$ boson from the jets 
originating from the tree-level
$\qq$ pair, whereas in the LRSM final states, one can instead reconstruct the masses of the heavy gauge bosons. 
Furthermore, the energy scales of the two models are largely separate. The energy scale of mTISM final states is set by the heavy neutrino mass, which, based on the LEP constraints~\cite{Abreu:1996pa,Adriani:1992pq}, is assumed to be greater than $100 \GeV$. Instead, the energy scale of LRSM final states is set by the masses of the heavy bosons, which, motivated by the earlier heavy neutrino searches, are assumed to be greater than $400 \GeV$. For these reasons, the event selection criteria are optimised separately for each model, although a common object selection is used in both cases.

\section{The ATLAS detector}

The ATLAS detector~\cite{ATLASmain} surrounds the interaction point and covers nearly the entire solid angle.
The detector consists of an inner detector (ID) tracking system, electromagnetic and hadronic calorimeters, and a muon spectrometer (MS) that surrounds the other detector systems. 
The ID tracking system consists of a silicon pixel detector, a silicon microstrip tracker, both covering $|\eta|<2.5$,~and a transition radiation
tracker covering $|\eta|<2.0$.
The ID tracker is immersed in a 2 T axial magnetic field provided by
a superconducting solenoid magnet. 
The electromagnetic accordion calorimeter is composed
of lead and liquid-argon (LAr) and provides coverage for $|\eta|<3.2$. Hadronic calorimetry is provided
by steel and scintillator tile calorimeters for $|\eta|<1.7$~and copper and LAr calorimeters for $1.5 < |\eta|<3.2$. Additional LAr calorimeters with copper and tungsten
absorbers cover the forward region. The MS consists of dedicated trigger chambers covering $|\eta|<2.4$~and precision tracking
detectors covering $|\eta|<2.7$. 
A system of three superconducting toroids (one in the barrel, two in the end-caps),~with eight coils each, provides the magnetic field for the MS.
A three-level trigger system is used to collect the data.
The first-level trigger (L1) is implemented in hardware, using a subset of detector information to reduce the event rate to no more than 75 kHz.
This is followed by two software-based trigger levels (L2 and EF), which together further reduce the event rate to less than 1 kHz.

\section{Background and signal simulation}
\label{sec-MC}

There are several SM interactions that can produce pairs of isolated charged leptons from vector boson decays, specifically from 
Drell--Yan and diboson processes and also from the
decay products of top quarks. These processes are modelled using events produced by Monte Carlo (MC) generators. A summary of the primary MC
generators used in this paper to model the background processes is presented in table~\ref{tab:MCinfo}. Processes that contribute to the background
with pairs of same-sign leptons are indicated by the label `SS' and processes which contribute with pairs of opposite-sign leptons, which are included in the charge-misidentification background estimate
(section~\ref{sec:cf}), are indicated by the label `OS'. The MC samples are normalised using the best available theoretical cross-sections, typically
next-to-leading order or next-to-next-to-leading order in QCD.

The production of top quark pairs and the production of a single top-quark in association with a $W$~boson
are simulated using {\sc MC@NLO 4.0.3}~\cite{1126-6708-2002-06-029,Frixione:2008yi} with the CT10 PDF set~\cite{CT10}. 
The production of single $Z$~boson and diboson ($W^\pm W^\mp$, $WZ$~and $ZZ$) events are simulated using {\sc Sherpa 1.4.1}~\cite{1126-6708-2009-02-007} with the CT10 PDF set. These samples include contributions
from virtual photons, with the requirement for electron pairs to satisfy $m_{e^+ e^-} > 0.1 \GeV$.
No requirement is made on $m_{\mu^+ \mu^-}$. The {\sc Sherpa} samples
 include leading-order matrix elements for the production of up to three additional
partons. The matching between the matrix elements and the parton shower is achieved using the CKKW method~\cite{1126-6708-2001-11-063}.
The process $qq \to q'q' W^{\pm}W^{\pm}$~is generated using {\sc Madgraph~2.1.1}~\cite{1126-6708-2003-02-027} with the MSTW 2008 PDF set~\cite{Martin:2009iq}.
The production of gauge bosons in association with top quark pairs is also simulated
with {\sc Madgraph~2.1.1} using the CTEQ6L1 PDF set.

\begin{table}[t]
\renewcommand{\arraystretch}{1.3} 
\centering
\small
\begin{tabular}{ c | c | c | c | c } 
Process & OS/SS & Generator & Parton shower & PDF set \\ \hline
\multicolumn{5}{l}{Background processes} \\
\hline
$Wt$  & \multirow{4}*{OS} & \multirow{2}*{\sc MC@NLO} & \multirow{2}*{\sc Herwig} & \multirow{6}*{CT10} \\ 
\cline{1-1}
$t\bar{t}$ & & & &  \\ 
\cline{1-1} \cline{3-4}
$Z$ & & \multirow{4}*{\sc Sherpa} & \multirow{4}*{\sc Sherpa} & \\ 
\cline{1-1}
$W^\pm W^\mp$ & & & & \\
\cline{1-2}
$WZ$ & \multirow{4}*{SS} & & & \\
\cline{1-1}
$ZZ$  & & & & \\
\cline{1-1} \cline{3-5}
$W^\pm W^\pm jj$ &  &  \multirow{2}*{\sc Madgraph} &  \multirow{2}*{\sc Pythia} & MSTW 2008\\ 
\cline{1-1} \cline{5-5}
$t\bar{t} + W/Z$ & & & & {CTEQ6L1}\\ 
\hline 
\multicolumn{5}{l}{Signal processes} \\
\hline
$W^{\pm} \rarr \lpm N$ & \multirow{3}*{SS}  & {\sc Alpgen} & \multirow{3}*{\sc Pythia} & CTEQ6L1 \\
\cline{1-1} \cline{3-3} \cline{5-5}
$W_{\text{R}}^{\pm} \rarr \lpm N$ & & \multirow{2}*{\sc Pythia} & & \multirow{2}*{MSTW 2008} \\
\cline{1-1}
$Z' \rarr N N$  & & & &  \\
\hline 
\end{tabular} 
\caption{Overview of primary MC samples used for the simulation of signal and background processes. The category labelled `OS/SS' refers to whether the process 
leads to pairs of opposite-sign (OS) or same-sign (SS) leptons. As described in section \ref{sec:cf}, OS MC samples are used in the prediction of the charge-misidentification background.}
\label{tab:MCinfo}
\end{table}

Table~\ref{tab:MCinfo} also shows how the various signal processes are modelled.
Signal events in the mTISM model are generated at leading order in QCD using the {\sc Alpgen 2.14} MC generator~\cite{ALPGEN,ALPGEN_HN} with the CTEQ6L1 PDF set. 
The events are generated for heavy Majorana neutrino masses between~$100$ and $500~\GeV$. Final states which contain exactly two prompt leptons \{$ee$, $\mu\mu$\}~with same-sign charge are produced.

The {\sc Pythia~8.170} generator~\cite{Sjostrand:2014zea} is used to generate LRSM events at leading order in QCD.
It is assumed in the model that the coupling between heavy gauge bosons and the heavy neutrino is equal
to the coupling between the respective SM gauge bosons and light neutrinos. 
Similarly, the couplings of the new gauge bosons and the quarks are assumed to be equal to the couplings between the 
SM gauge bosons and the quarks.
In the {\sc Pythia} implementation of this process,
all of the decay products of heavy neutrinos are distributed isotropically and so the heavy neutrino decays 
are independent of $\mWR$, with the assumption that $\mWR > \mN$.
In this approximation, the decay of the $\ZR$~boson is also independent of $\mWR$.
The events are generated with $\WR$~boson masses between
$0.6 \TeV$ and $4.5 \TeV$ and $\ZR$~boson masses between $0.4 \TeV$ and $3.6 \TeV$. At each $\ZR$~and $\WR$~mass point, the heavy neutrino mass is varied upward from $50 \GeV$ to
at most $100 \GeV$ below the mass of the heavy gauge boson. 
At each mass point, a sample is generated assuming no mixing between the heavy neutrinos, which results in final states containing same-flavour leptons \{$ee$, $\mu\mu$, $\tau\tau$\}.

Parton showering, fragmentation, hadronisation and the modelling of the underlying event for all
{\sc Madgraph} and {\sc Alpgen} samples are performed with {\sc Pythia 8.165} and for {\sc MC@NLO} samples with {\sc Herwig 6.520}~\cite{1126-6708-2001-01-010} and {\sc Jimmy 4.31}~\cite{Butterworth:1996zw}. 

The effect of multiple $pp$ collisions in the same or different bunch crossings is incorporated into the simulation by overlaying minimum-bias events
generated using {\sc Pythia 8}
onto hard-scatter events, where the number of additional interactions is distributed in the same
way as in data.
All the background samples are produced using a simulation of the ATLAS detector~\cite{sim10} based on {\sc Geant4}~\cite{Agostinelli2003250}. 
The signal samples are processed through a fast simulation using a parameterisation of the performance of the ATLAS electromagnetic and hadronic calorimeters~\cite{1742-6596-331-3-032053}, and {\sc Geant4} 
in the ID and MS. 
Both the signal and background samples are then processed with the same
reconstruction software as the data. 
Small differences between data and MC simulation in the lepton reconstruction, identification and trigger efficiencies are corrected for by using specific data-driven measurements.

\section{Data sample and event selection}
\label{evsel}

The events used were selected from $pp$~collision data with an integrated luminosity of 20.3 $\ifb$ collected by ATLAS in $2012$.  
Quality criteria are applied to suppress non-collision backgrounds such as cosmic-ray muons, beam-related backgrounds, and spurious noise in the calorimeters. 

\subsection{Object reconstruction and selection}
\label{sec:obj}

The search uses reconstructed electrons, muons, jets and a measurement of the missing transverse momentum. 

Electrons are required to satisfy tight identification requirements~\cite{Egamma_ID} and to have $\pt~> 20$~GeV and $|\eta|<2.47$.
Any electron in the transition region between the barrel and end-cap calorimeters ($1.37 < |\eta| < 1.52$) is rejected. 
In order to avoid double counting electrons as jets, the nearest jet within $\Delta R(e,\mathrm{jet}) = 0.2$ of an electron and with
$\pT~<~2\et$, where $\et = E \sin \theta$~is the transverse energy deposited by the electron, is rejected.

Muons are required to be reconstructed in the MS and successfully matched to a good-quality track in the ID~\cite{muonsys}. It is required that muons have $\pt > 20 \GeV$ 
and $| \eta | < 2.5$.
In order to suppress muons with misidentified charge, it is required that
there is a consistent measurement of charge in the MS and ID. Muons with $\pt < 80$~GeV are required to be well
separated from jets, such that $\Delta R(\mu, \mathrm{jet}) > 0.4$, where $\Delta R = \sqrt{(\Delta\eta)^2 + (\Delta\phi)^2}$.  

Jets are reconstructed using the anti-$k_{t}$ clustering algorithm~\cite{Cacciari200657,Cacciari:2008gp} with the radius
parameter set to $0.4$. Jets are calibrated~\cite{Aad:1705576,jes_unc} using an energy- and $\eta$-dependent simulation-based calibration
scheme, with in-situ corrections based on data. The impact of multiple overlapping $pp$
interactions is accounted for using a technique that provides an event-by-event and jet-by-jet
correction~\cite{Cacciari2008119}. Events are rejected if any jet is identified as originating from beam-halo effects or calorimeter noise.
Jets are required to have $\pt > 20$~GeV and $|\eta|<2.8$. The $\pt$~requirement is chosen in order to maximise the acceptance for the mTISM model.
For jets with $\pt < 50$~GeV within the acceptance of the tracking detector ($|\eta| < 2.4$), the
 `jet vertex fraction' (JVF)~\cite{Miller:1206864} is required to be greater than 0.5. 
 The jet vertex fraction is calculated by summing the $\pt$ of tracks associated with the jet and matched to the selected primary vertex, and dividing it by the sum of the $\pt$ of all tracks associated with the jet.
 
The primary vertex of the event is defined as the reconstructed vertex with the highest $\sum \pt^2$, consistent with the beam spot position, where the sum is over all tracks associated with the candidate primary vertex.

The missing transverse momentum, $\MET$, is used to identify invisible particles such as light neutrinos that escape detection.
The $\MET$ quantity is calculated as the magnitude of the negative vector sum of all reconstructed particles momenta, 
including muons, electrons,
photons, and jets, as well as clusters of calorimeter cells, not associated with these objects.

\subsection{Lepton isolation criteria}
\label{sec:lepiso}

Backgrounds due to misidentified leptons and non-prompt leptons, which are described in detail in section~\ref{sec:fake}, can be suppressed by requiring that leptons are isolated from
other activity in the event.
Because of the different background compositions for electrons compared to muons and of the different response of the detector to isolated electrons and muons from prompt sources, different
isolation criteria are used for the two lepton flavours. 

Electrons are required to satisfy $\left( \pt^{\text{C}2} + 1\ \mathrm{GeV}\right)/\et < 0.05$~and
$\et^{\text{C}3}/\et < 0.05$, where
$\pt^{\text{C}2}$~is the sum of the $\pt$ of all tracks within a cone of $\Delta R=0.2$\ 
around the electron, excluding the electron track itself, and  $\et^{\text{C}3}$~is the sum of the transverse energy in a cone of $\Delta R=0.3$~around the electron,
excluding the electron itself.
The criteria are looser at high electron
$\et$~in order to maintain high efficiency.

Muons with $\pt < 80 \GeV$ are required to have $\pt^{\text{C}3}/\pt < 0.05$~and $\et^{\text{C}2}/\pt < 0.05$, 
where $\pt^{\text{C}3}$~is the sum of the $\pt$ of all tracks within a cone of $\Delta R=0.3$\ 
around the muon, excluding the muon track itself, and $\et^{\text{C}2}$~is the sum of transverse energy measured in the calorimeter within a cone of $\Delta R=0.2$~around the muon,
excluding energy deposits associated with the muon track.
For muons with $\pt > 80$~GeV the requirements are relaxed in order to maintain high efficiency. 
Muons with $\pt > 80$~GeV are either required
to satisfy $\et^{\text{C}2}/\pt < 0.05$, or if they are within $\Delta R=0.4$~of a jet they can additionally be selected if
$(m_{\mu j} - m_{j}) > 10$~GeV, where $m_{j}$~is the reconstructed mass of the jet closest to the muon and $m_{\mu j}$~is the invariant mass of the jet and the muon.
The latter criterion is efficient for the decay of a boosted heavy neutrino decaying into a muon and a $\qq$~pair, while rejecting a large fraction of misidentified muons.

Both the muons and electrons must satisfy a set of requirements on the impact parameters at the primary vertex in order to further suppress leptons originating from heavy-flavour decays. They are required to have a transverse impact parameter, $d_{0}$,
which satisfies $|d_{0}| < 0.2 \ \rm{mm}$ and $|d_{0}|/\sigma(d_{0}) < 3$, where $\sigma(d_{0})$ is the uncertainty on $d_{0}$. It is also required that the product 
of the longitudinal impact parameter ($z_0$) and the sine of the polar angle of the lepton ($\theta$) satisfy $|z_{0}\sin\theta| < 2 \ \rm{mm}$. 

\subsection{General event selection}
\label{sec:gensel}

The events are required to satisfy one of a suite of triggers~\cite{6551337} that select events with either one or two high-$\pt$~leptons. 
This analysis uses single-lepton ($e$ or $\mu$) triggers with a 24 GeV $\pt$ threshold. 
The analysis also uses a dimuon trigger for events in which one muon has satisfied a $\pt$ threshold of 20 GeV and a second muon has satisfied a threshold of 8 GeV. 
The choice of triggers is found to maintain the highest possible signal efficiency in each channel across the presented range of heavy neutrino masses. The electron (muon) trigger efficiencies for offline selected electrons (muons) are $\gtrsim  94\% \ (70\%)$ and $\gtrsim 85\% \ (90\%)$ in the barrel and end-cap, respectively. The total efficiency
for a single electron (muon) within the detector acceptance to satisfy the full lepton selection described in sections~\ref{sec:obj} and~\ref{sec:lepiso}, including the trigger requirement, is approximately $54\% \ (70\%)$.

The highest-$\pt$ lepton in an event is required to satisfy $\pt > 25 \GeV$. The choice of lepton $\pt$ threshold is
dictated by the trigger requirements.
It is required that at least one lepton with $\pt > 25 \GeV$ is matched to one of the described triggers. In the case of the event being selected by the dimuon
trigger, two muons must be matched to the trigger.  
Any other leptons
must satisfy $\pt > 20 \GeV$.

Events are required to contain exactly two leptons from $\{ee, \mu\mu\}$ with same-sign charge, where 
the two leptons must have ID tracks associated with the same vertex.
To remove the small background arising from muon bremsstrahlung in the ID or in the first layers of the EM
calorimeter, events are rejected if a muon's ID track is also reconstructed as an electron.
Backgrounds from $WZ$ and $ZZ$ decays are suppressed by rejecting events which contain an additional lepton, where the additional lepton is selected with looser identification requirements and no requirements on the isolation variables. The impact of the latter criterion on the signal efficiency is negligible compared to the overall uncertainty on the signal acceptance.

\subsection{Selection criteria for mTISM signal events}
\label{sec:mTISM_sigsel}

The signal region for the mTISM model is defined for events containing, in addition to the two leptons, at least two jets. The invariant mass, $m_{jj}$, 
of the two highest-$\pt$ jets is required to lie in the range $60 < m_{jj} < 100 \GeV$. This
selects events consistent with an on-shell $W$ boson decaying to a $\qq$ pair. The invariant mass of the two leptons ($m_{\ell\ell}$) 
is required
to be greater than $40 \GeV$, a selection that has high efficiency for the signal over the full range of $m_N$~values considered.
In the $ee$~channel, the charge-misidentification background (described in section~\ref{sec:cf}) is suppressed by requiring that the invariant mass of the two leptons is
outside a window around
the $Z$ boson mass, $|m_{\ell\ell} - m_Z| > 20 \GeV$. Backgrounds due to electroweak processes producing same-sign leptons
are dominated by those including at least one light neutrino, particularly those arising from diboson production processes. Such backgrounds may have high $\MET$ compared to the mTISM signal, so the events in the signal region are required to have $\MET < 40 \GeV$.

The total efficiency (including the detector acceptance) for signal events to satisfy all selection criteria is lower in $ee$ events than in $\mu\mu$ events as electrons have lower efficiency to satisfy the identification and isolation
criteria. The efficiency for leptons in mTISM signal events to satisfy the object selection increases as a function of the lepton $\pt$. The total efficiency therefore increases as a function of $m_N$, from approximately 0.5\% to 24\% in the $ee$ channel and 3\% to 30\% in the $\mu\mu$ channel.

\subsection{Selection criteria for LRSM $\WR$ and $\ZR$ signal events}
\label{sec:LRSM_sigsel}

The signal region for heavy neutrinos produced in the decays of $W_{\text{R}}$ bosons is defined for events containing at least one jet. 
To exploit the high energy scale of the signal events it is required that 
$m_{\ell\ell} > 110 \GeV$ and the invariant mass of the system consisting of the two leptons and one or two jets must satisfy $m_{\ell \ell j(j)} > 400 \GeV$.
If the event contains more than two jets, the two highest-$\pt$~jets are used.

The signal region for heavy neutrinos produced in the decays of $\ZR$ bosons uses the same requirement on the dilepton mass, but requires at least two jets
and the invariant mass of the system consisting of the two leptons and two to four jets must satisfy $m_{\ell \ell jj(jj)} > 200 \GeV$.
If the event contains more than four jets, the four highest-$\pt$~jets are used. 
A lower invariant mass of the system of $200$~GeV is considered in the search for $\ZR$ bosons compared to the value of $400$~GeV used for $W_{\text{R}}$ bosons since previous searches for heavy neutrinos have
 already set strong constraints for $\mWR < 400$~GeV~\cite{ATLAS_LRSM}, whereas there are no such limits for $\ZR$ production.

Allowing only one jet in the $W_{\text{R}}$ decay or two jets in the $\ZR$ decay (rather than two and four respectively) increases the signal efficiency for the case when the heavy neutrino is boosted ($\mVR \gg \mN$) 
so that the $\qq$ pair produced in the decay of the off-shell $W_{\text{R}}$ results in a single jet reconstructed in the detector. For $\mVR \gg \mN$, up to 60\% of events contain 
only one jet in $W_{\text{R}}$\ events or two jets in the case of $\ZR$\ decays. The total selection efficiency for LRSM events depends on
the masses $m_{V_{\text{R}}}$ and $m_N$ and also the ratio $m_{V_{\text{R}}}/m_N$. The total efficiency (including the detector acceptance) ranges from approximately $0.5\%$ to $25\%$ in the $ee$ channel, and from approximately $1.5\%$ to $30\%$ in the $\mu\mu$ channel.
For small values of $\mN / \mVR$, the efficiency decreases rapidly and is below $15\%$~for $\mN / \mVR < 0.1$,~because the heavy neutrino decay products are highly boosted and the leptons are less isolated.

\section{Background estimation}
\label{backest}

The background is evaluated in the signal regions according to three categories of lepton pairs.
The first two categories describe the contribution due to leptons
originating from SM processes that produce prompt isolated leptons.
The first category, labelled as `prompt' in the following and discussed in section~\ref{sec-prompt}, corresponds to 
irreducible background from true same-sign prompt lepton pairs. The second category, labelled as `charge-flip' in 
the following and discussed in section~\ref{sec:cf}, corresponds to true opposite-sign prompt lepton pairs, in which 
one lepton has its charge mismeasured. 
The third category, labelled as `non-prompt' in the following and discussed in section~\ref{sec:fake}, corresponds to one or 
both leptons being either a non-prompt lepton from semileptonic heavy-flavour decays,
a jet misidentified as a lepton or, in the case of electrons, photons misidentified as leptons.
The first category of background events are entirely estimated using the SS MC samples described in section~\ref{sec-MC},
the second category is estimated using OS MC samples in conjunction with a measurement in the data of the charge
misidentification rate and the third category is estimated from data.

\subsection{Background from prompt same-sign leptons}
\label{sec-prompt}

The background from SM processes that lead to two same-sign prompt leptons is referred to as the prompt background and is estimated using the MC
samples described in section~\ref{sec-MC}.
The largest contribution to the SM background in the signal regions originates from $WZ$~and $ZZ$~events.
Other prompt contributions were estimated to be negligible; these include processes involving the production of three electroweak gauge bosons or of a Higgs boson.
The simulation of the $WZ$~and $ZZ$~backgrounds is validated by selecting events with either three or four charged leptons satisfying the selection cuts described in section~\ref{sec:gensel}. These events arise predominantly from $WZ$~and $ZZ$~production, respectively, with a negligible contribution expected from the signal processes. The expected and observed jet multiplicity and leading jet $\pt$ distributions for these events are shown in figure~\ref{fig:dbmc}. The number of events predicted by the simulation
is found to be in good agreement with the data and the kinematic properties of the events are adequately described by the simulation. \\

\begin{figure*}[bthp]
\centering
\subfigure[]{\includegraphics[width=0.49\textwidth]{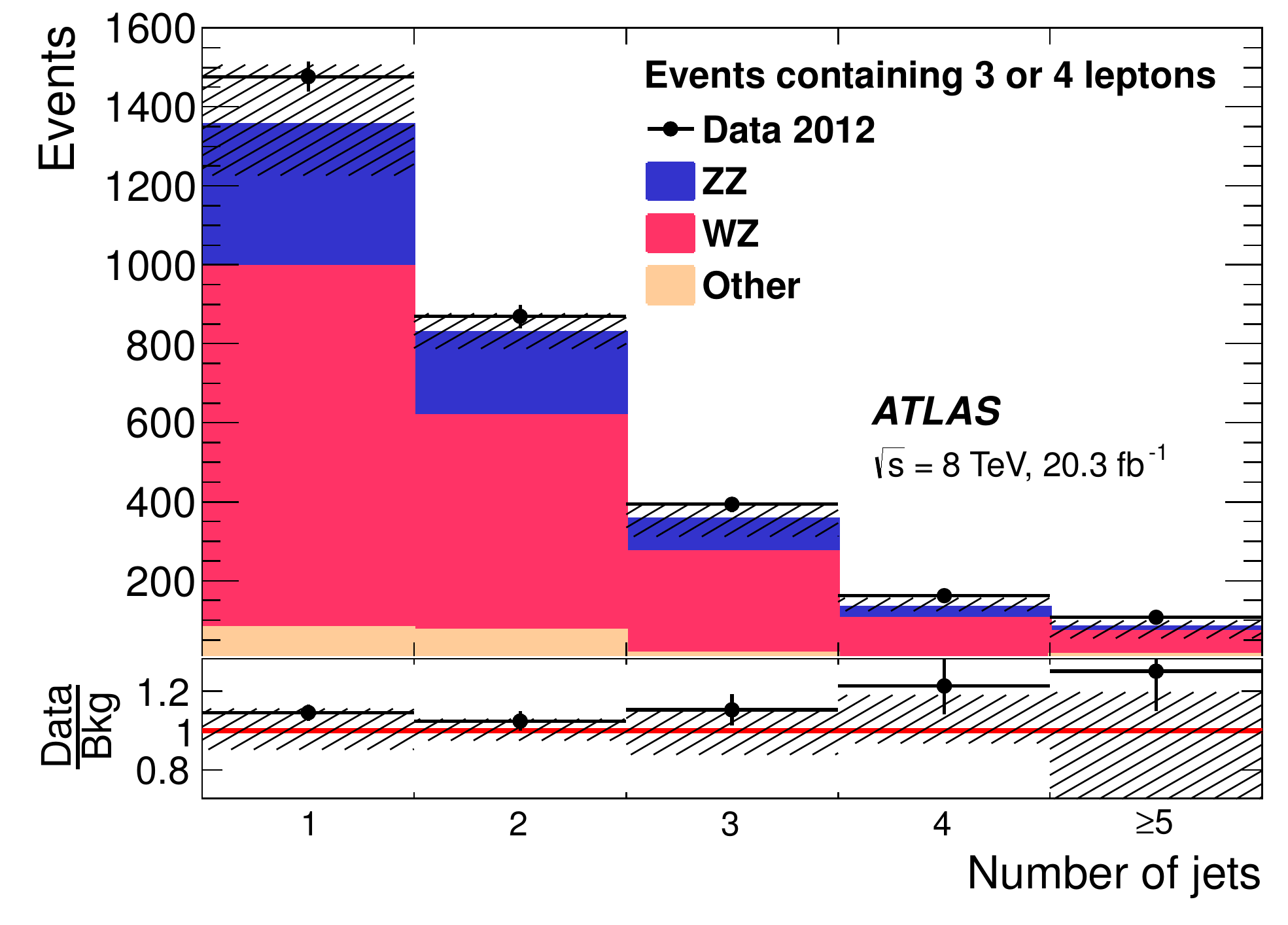}}
\subfigure[]{\includegraphics[width=0.49\textwidth]{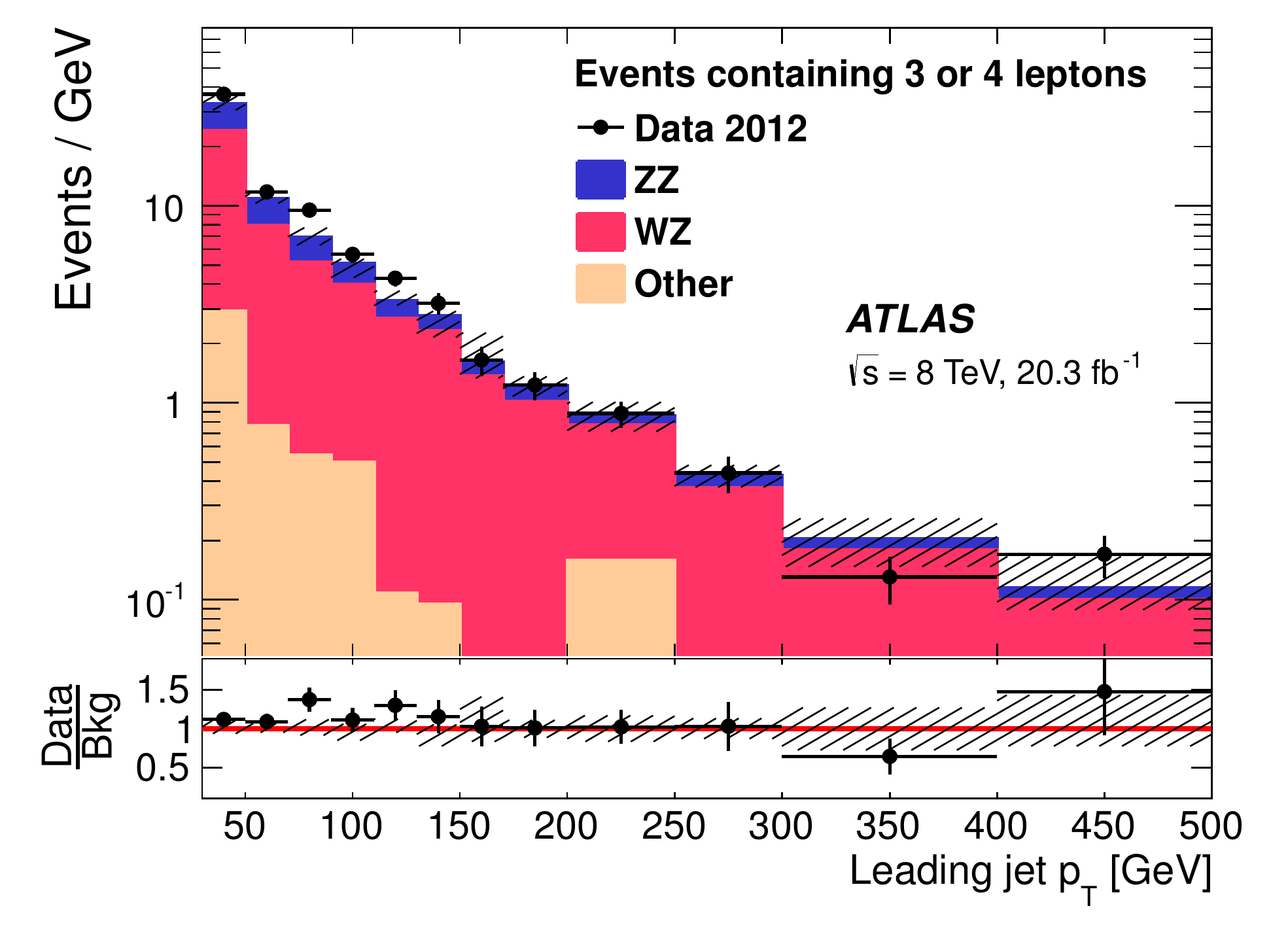}}
  \caption{Distribution of (a) the number of jets and (b) the leading jet $\pt$ in events containing any combination of exactly three or four leptons. 
  The events must contain one lepton with $\pt > 25$~GeV and all other leptons must satisfy  $\pt > 20$~GeV. The contribution labelled `Other'
  is from processes described in section~\ref{sec-prompt} (with MC samples described in section \ref{sec-MC}), other than the 
  contributions from $WZ$ and $ZZ$, which are labelled separately.
 The shaded bands indicate the experimental uncertainties on the total expected background, including all contributions described in section \ref{sec:systMC}, but not including any uncertainty on the $WZ$~and $ZZ$~cross-sections. The lower plots show the ratio of data to the total expected background.}
  \label{fig:dbmc}
\end{figure*}

\subsection{Background from prompt opposite-sign leptons}
\label{sec:cf}

SM processes that produce opposite-sign leptons can also enter into the signal regions if the charge of one lepton is incorrectly measured in the detector. This is referred
to as the `charge-flip' background and includes pairs of opposite sign-leptons produced in $\ttbar$, $W^\pm W^\mp$ and $Z$ processes. The MC samples used to model
these processes are described in section~\ref{sec-MC}.

The probability for a charge-flip event to occur is measured using $Z\to\mu\mu$~and $Z\to ee$~events, which are identified using events, which have exactly two leptons
with an invariant mass close to the mass of the $Z$ boson.
For muons, the charge-flip rate is estimated by comparing the two independent measurements of the muon charge in the MS and the ID. The charge-flip rate is found, as expected from simulation,
to be consistent with zero, and contributes a negligible number of events to the expected background.

There is however a sizeable charge-flip rate for electrons, due to bremsstrahlung photons produced in the ID and subsequently converting to electron pairs. A sample of $Z\to ee$~events in data is used to perform a single maximum-likelihood fit, which extracts the electron charge-flip rate as a function of $\eta$~\cite{flip}. The contribution to these events from fake and non-prompt electrons (see section~\ref{sec:fake}) is subtracted from the data sample prior to the fit.
The measured rate, which is strongly correlated to the amount of material traversed by the electron in the ID, is found to be approximately $10^{-4}$~for electrons within the barrel region ($|\eta|<1.0$), increasing to $10^{-2}$~at the edge of the detector acceptance ($|\eta|=2.47$).
The data-to-MC ratio of the measured electron charge-flip rate is used as a correction factor to the charge-flip rate obtained in the simulation. The $\pt$-dependence of the charge-flip rate is therefore directly taken from the MC simulation.

\subsection{Background from fake and non-prompt leptons}
\label{sec:fake}

Events where jets or photons are misidentified as leptons (`fakes') or events
with non-prompt leptons which originate from semileptonic heavy-flavour 
decays constitute a significant background, which is referred to as the non-prompt background.
These processes include $W$~+~jets and $t\bar{t}$~production, where one lepton originates from a vector boson decay and the other lepton
is misidentified or from a non-prompt decay.
This background cannot be reliably predicted from MC simulation and is estimated directly in each of the signal regions, using the data-driven matrix method~\cite{topxs}.
The matrix-method characterises leptons which satisfy `loose' identification criteria as being from prompt or fake/non-prompt sources according to their probabilities
to satisfy the full lepton identification criteria. 

\clearpage

A `loose' electron is defined with an identification requirement that is relaxed from the `tight' to `medium' operating point compared to the
standard electron selection described in section~\ref{sec:obj}. 
The selection criteria for a `loose' muon are identical to the full selection described in section~\ref{sec:obj}, with the same requirements on the impact parameters
described in section~\ref{sec:lepiso} but without a requirement to satisfy any isolation criteria. 

The measurement of the probabilities $r$ and $f$ for `loose' leptons from prompt or fake/non-prompt sources respectively
 to satisfy the full lepton
identification criteria are the key factors in the estimate of the non-prompt backgrounds.
The probabilities $r$ are measured in $Z \rightarrow \ell \ell$ events as these events are dominated by prompt leptons. 

The probabilities $f$ are measured in a selection of events which contain a large number of fake and non-prompt leptons. In these control samples,
any residual prompt background is subtracted using MC estimates, and low-mass hadronic resonances are excluded
by requiring $m_{\ell\ell} > 15 \GeV$.
For electrons, the probabilities are measured in events with at least one jet and with exactly one electron. 
To suppress events containing $W$~decays it is required that $|\Delta\phi(e, \met)| < 0.5$ and $\met + m_{\text{T}} < 40$~GeV, where $m_T$ is the transverse mass.\footnote{$m_{\text{T}} = \sqrt{2p_{\text{T}}E_{\text{T}}^{\mathrm{miss}}(1-\cos\left(\Delta\phi\left(e,E_{\text{T}}^{\mathrm{miss}}\right))\right)}$, where $p_{\text{T}}$ refers to the electron.} The probabilities for muons are 
measured in events containing 
pairs of muons chosen with transverse impact parameter requirements which differ from the standard selection, 
such that $|d_{0}| < 10$~mm and
$|d_{0}/\sigma(d_{0})| > 5$. This sample of pairs of muons
is expected to have a composition similar to the signal region, making it suitable for the measurement of $f$.
A correction factor is applied to $f$ for muons to account
 for the fact that muons with high $|d_{0}/\sigma(d_{0})|$ have a lower probability than muons with low $|d_{0}/\sigma(d_{0})|$ to satisfy the isolation criteria.
 The correction factor of approximately 1.4 is measured using $b\bar{b}$ and $t\bar{t}$ MC simulation.

A validation region is defined for same-sign $\mu\mu$ events which have satisfied the dimuon trigger (section \ref{sec:gensel}) with one muon satisfying $\pt > 25 \GeV$
and the second satisfying $10 \GeV < \pt <  20 \GeV$.  As the $\pt$ of the second muon is relatively low,
this region is dominated by non-prompt background events. 
The jet multiplicity measured in events in this region is compared to the expected background as shown in figure \ref{fig:sscr}. The overall level of agreement is within one
standard deviation for up to five jets in the event.

 \begin{figure}[bth]
 \centering
  \includegraphics[width=0.65\textwidth]{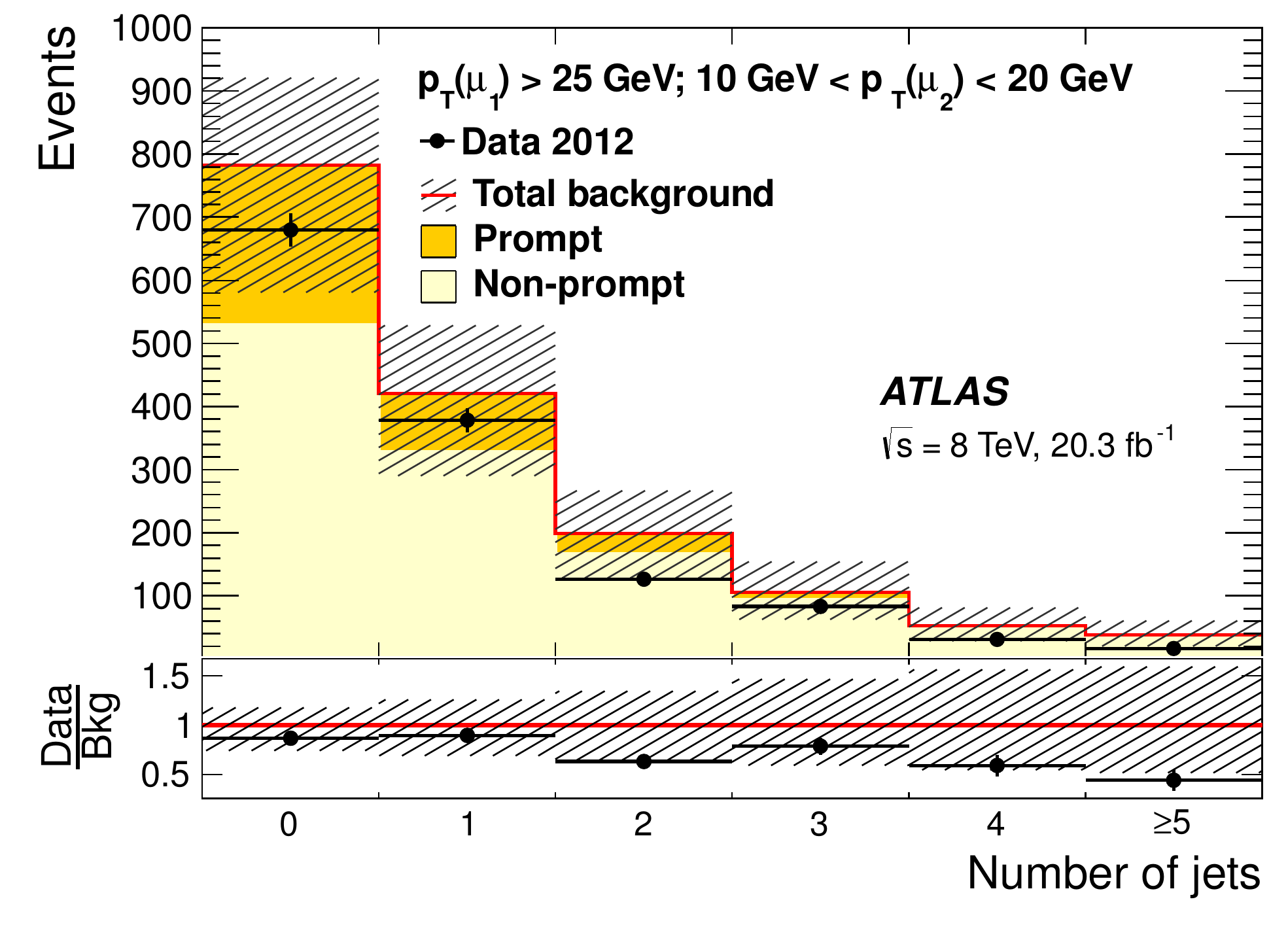}
  \caption{Distribution of the number of jets in a validation region consisting of events containing exactly two same-sign muons 
  with one muon satisfying $\pt > 25 \GeV$ and the second satisfying $10 \GeV < \pt <  20 \GeV$.
 The shaded bands indicate the total uncertainty, including all contributions described in section \ref{sys}, on the total expected background and
  the lower plots show the ratio of data to the total expected background.}
  \label{fig:sscr}
\end{figure}

\subsection{Validation of background estimates}

The validation of prompt, charge-flip and non-prompt background estimates is considered in this section.

The combined background estimate can be evaluated using events containing exactly two
same-sign leptons and no jets. This sample of events is orthogonal to each of the different signal regions described in 
sections~\ref{sec:mTISM_sigsel} and~\ref{sec:LRSM_sigsel}, and is expected to contain only a negligible contribution from possible signal events.
 Comparisons of the distributions of the $\met$ and lepton $\pt$ as measured in data and estimated from the background
 predictions from the three sources described above are shown in figure~\ref{fig:0jetcr}. 
The overall agreement between data and background prediction is within approximately one standard deviation in both the $ee$~and $\mu\mu$~channels.

An additional sample of events is considered, which fully includes all signal regions. This region is defined for events containing exactly two same-sign leptons,
with no requirement on the number of jets in the event. The contamination from signal events in this sample is less than 2\% of the total background in both channels.
The distributions of the leading jet $\pt$ and the number of jets in these events, in the $ee$ and $\mu\mu$ channels, as measured in data and estimated from the background predictions
 from the three sources described above, are shown are shown in figure~\ref{fig:0pjetcr}.
The overall agreement between data and prediction in the two channels is within approximately one standard deviations. \\
 
  \begin{figure}[ht]
\centering
\subfigure[]{\includegraphics[width=0.49\textwidth]{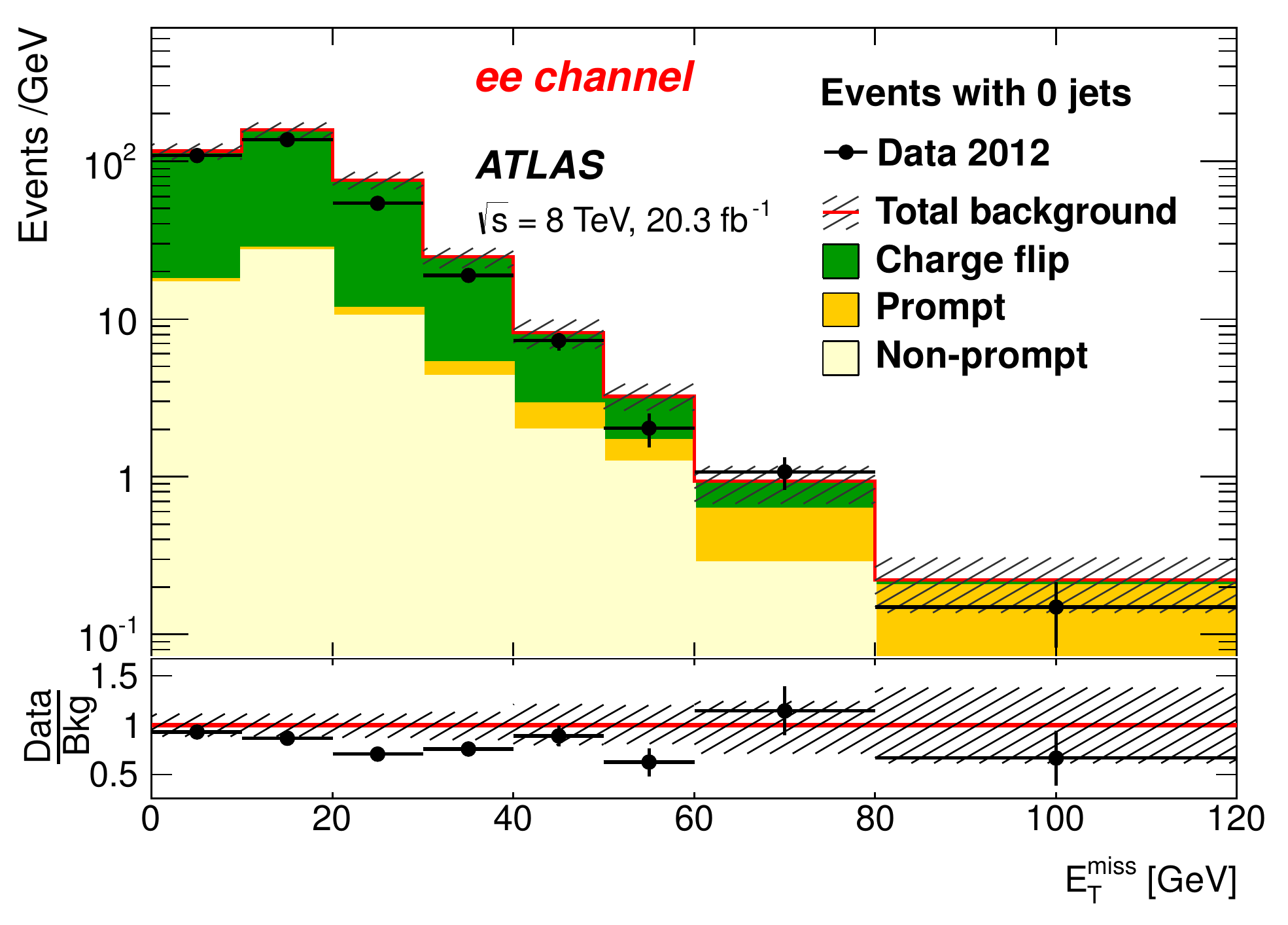}}
\subfigure[]{\includegraphics[width=0.49\textwidth]{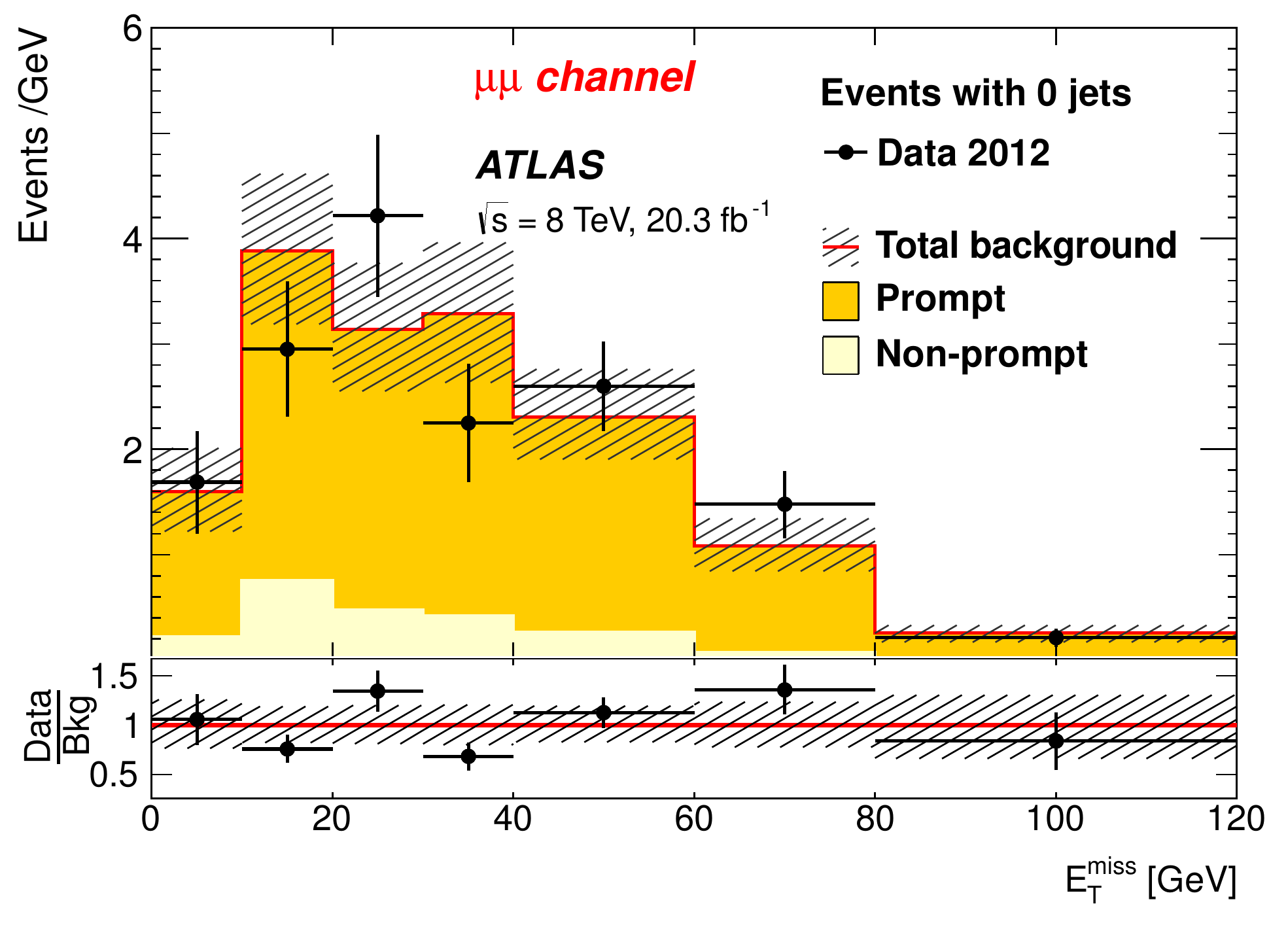}}
\subfigure[]{\includegraphics[width=0.49\textwidth]{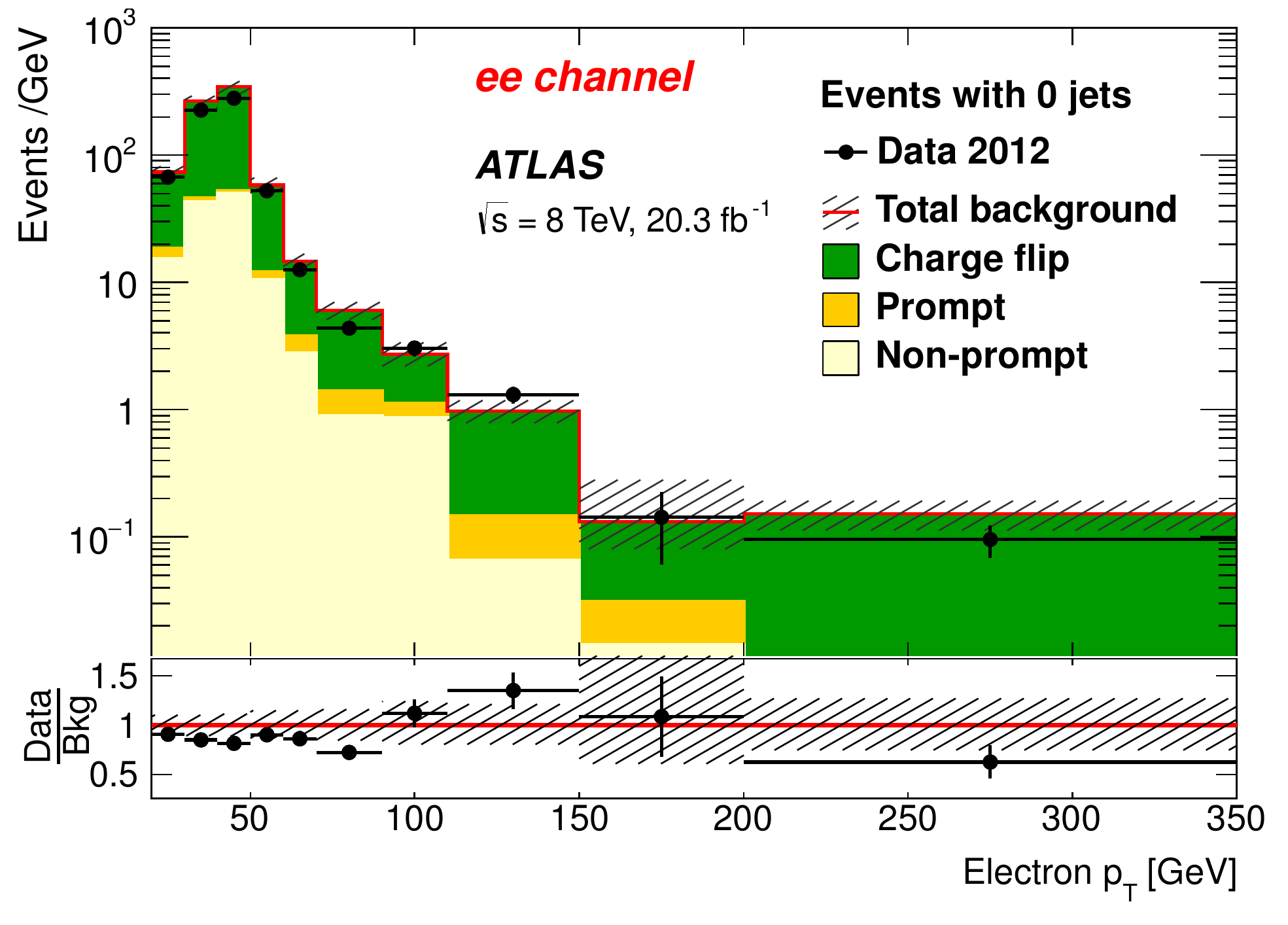}}
\subfigure[]{\includegraphics[width=0.49\textwidth]{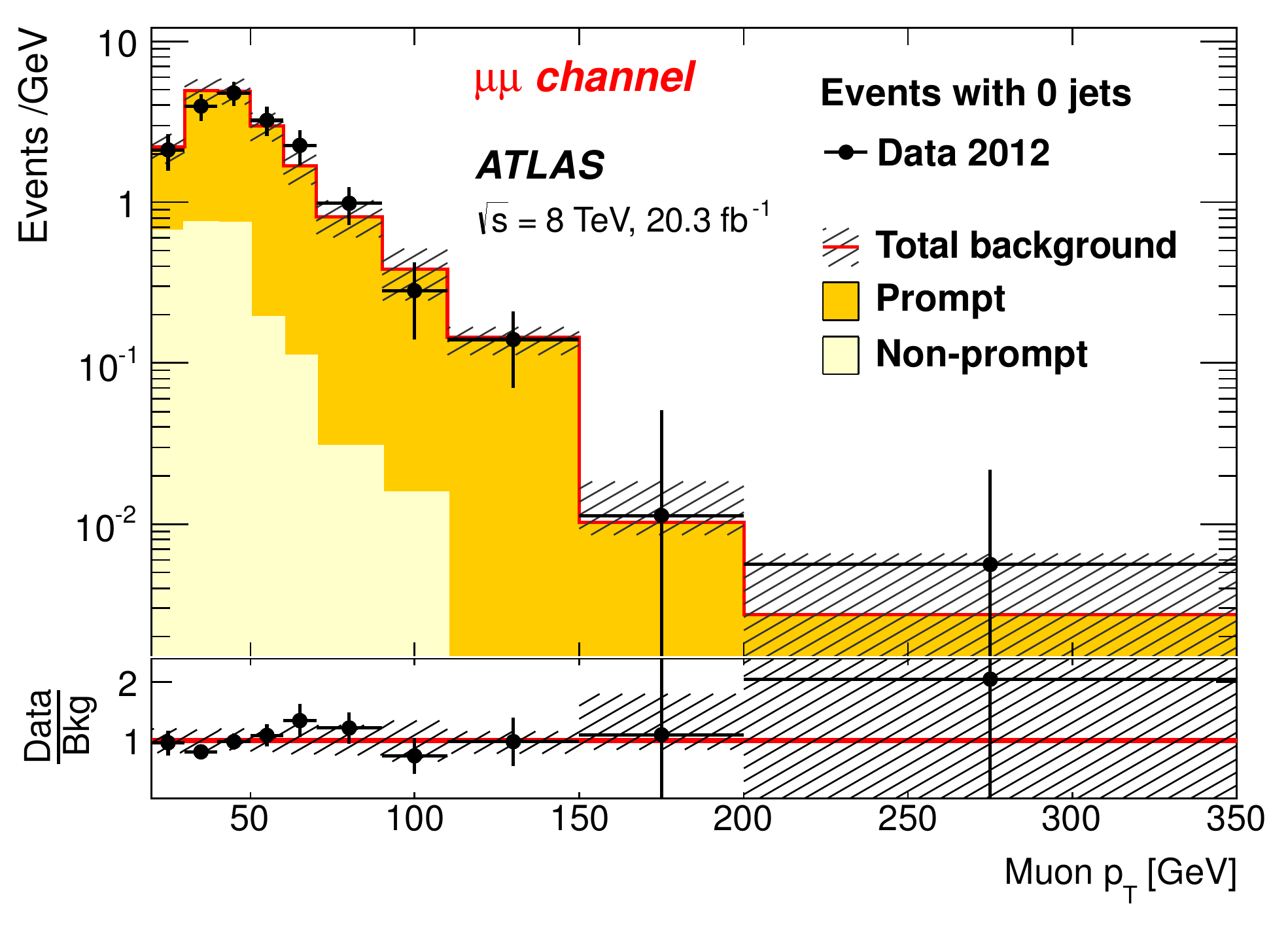}}
  \caption{The $\met$ (top) and lepton $\pt$ (bottom) distributions for the $ee$ (left) and $\mu\mu$ (right) channels in a validation region consisting of events with exactly two same-sign leptons and no jets.
The shaded bands indicate the total uncertainty, including all contributions described in section \ref{sys}, on the total expected background and
  the lower plots show the ratio of data to the total expected background.}
 \label{fig:0jetcr}
\end{figure}

\clearpage

 \begin{figure}[p]
\centering
\subfigure[]{\includegraphics[width=0.49\textwidth]{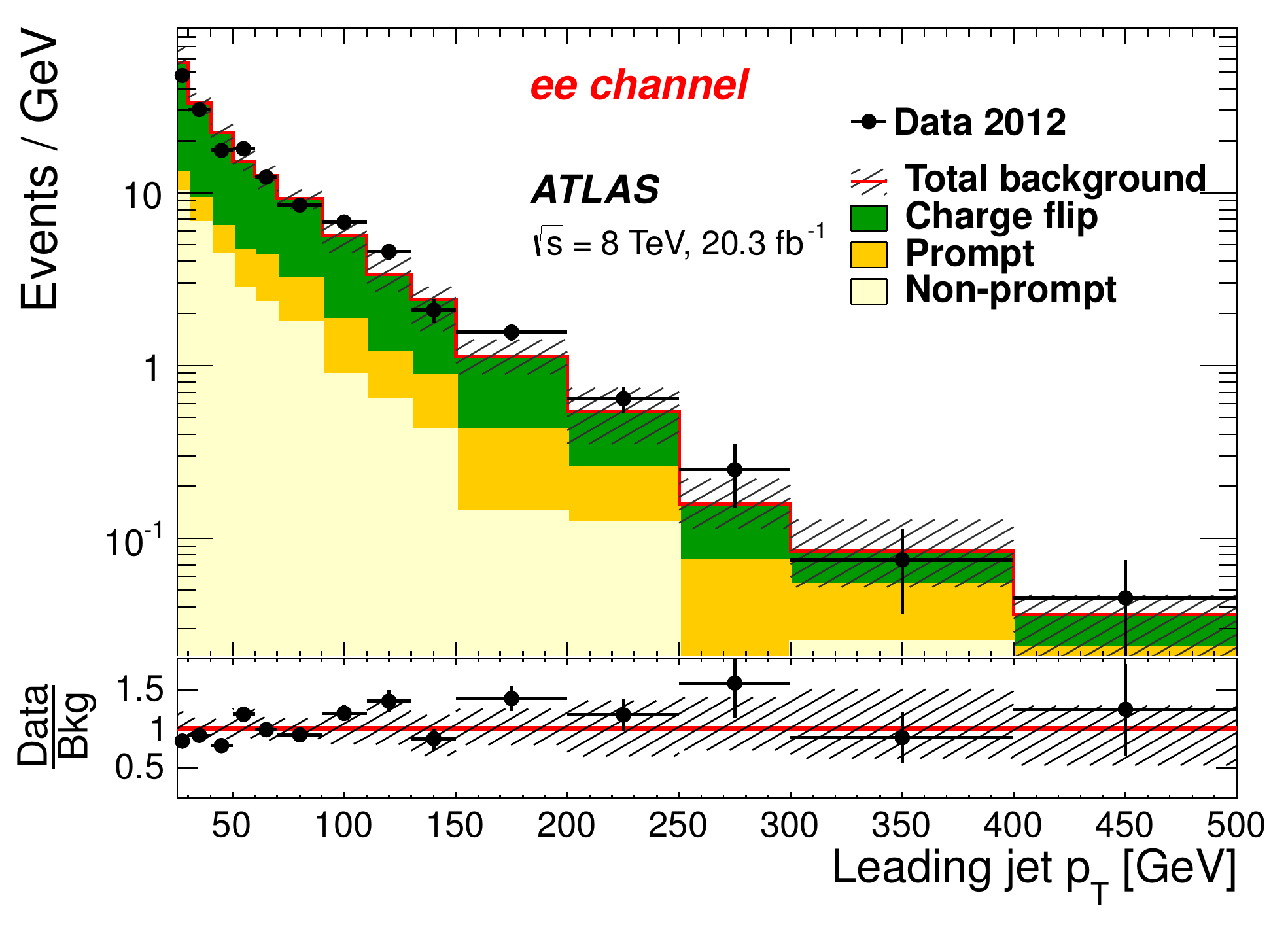}}
\subfigure[]{\includegraphics[width=0.49\textwidth]{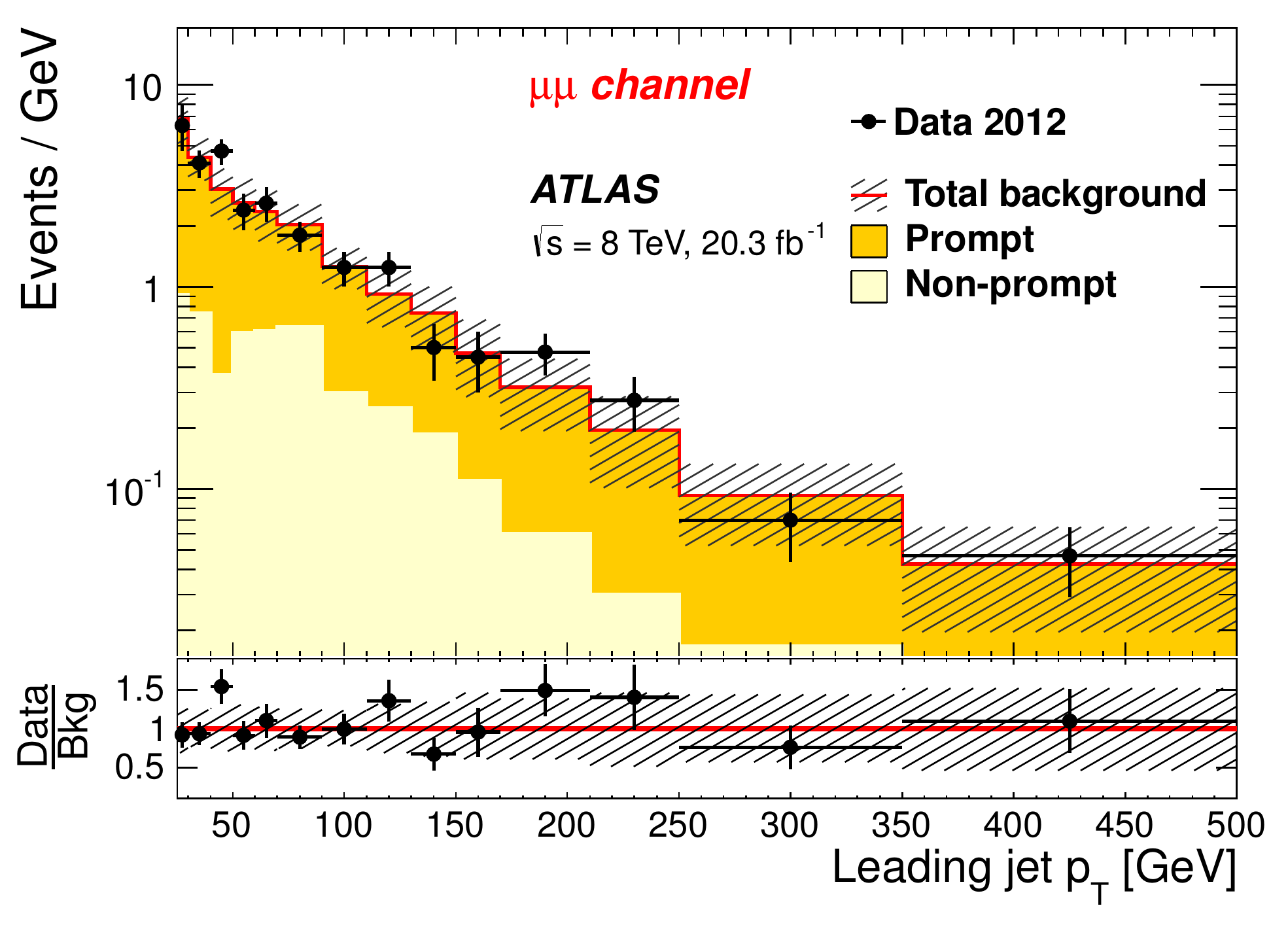}}
\subfigure[]{\includegraphics[width=0.49\textwidth]{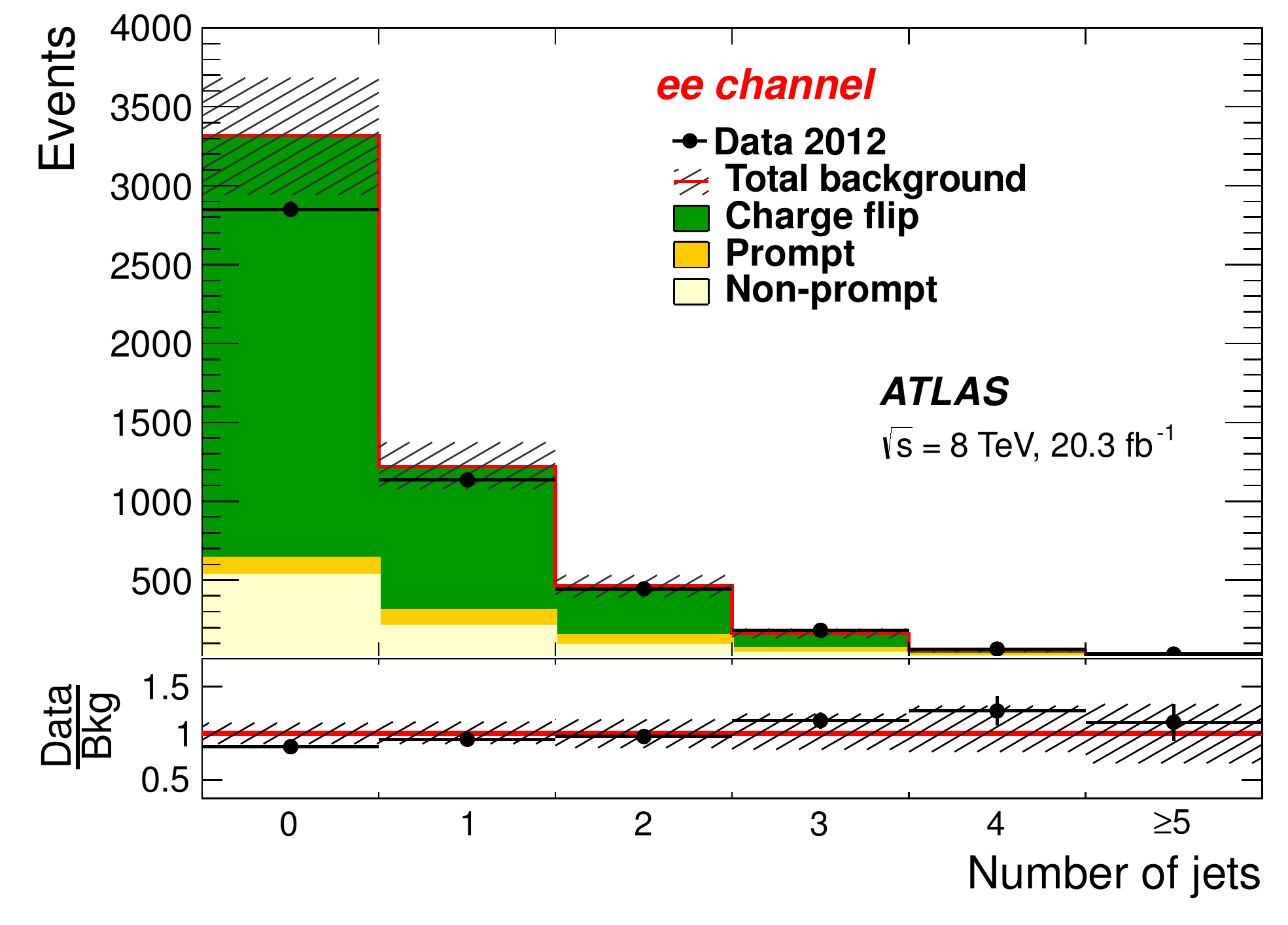}}
\subfigure[]{\includegraphics[width=0.49\textwidth]{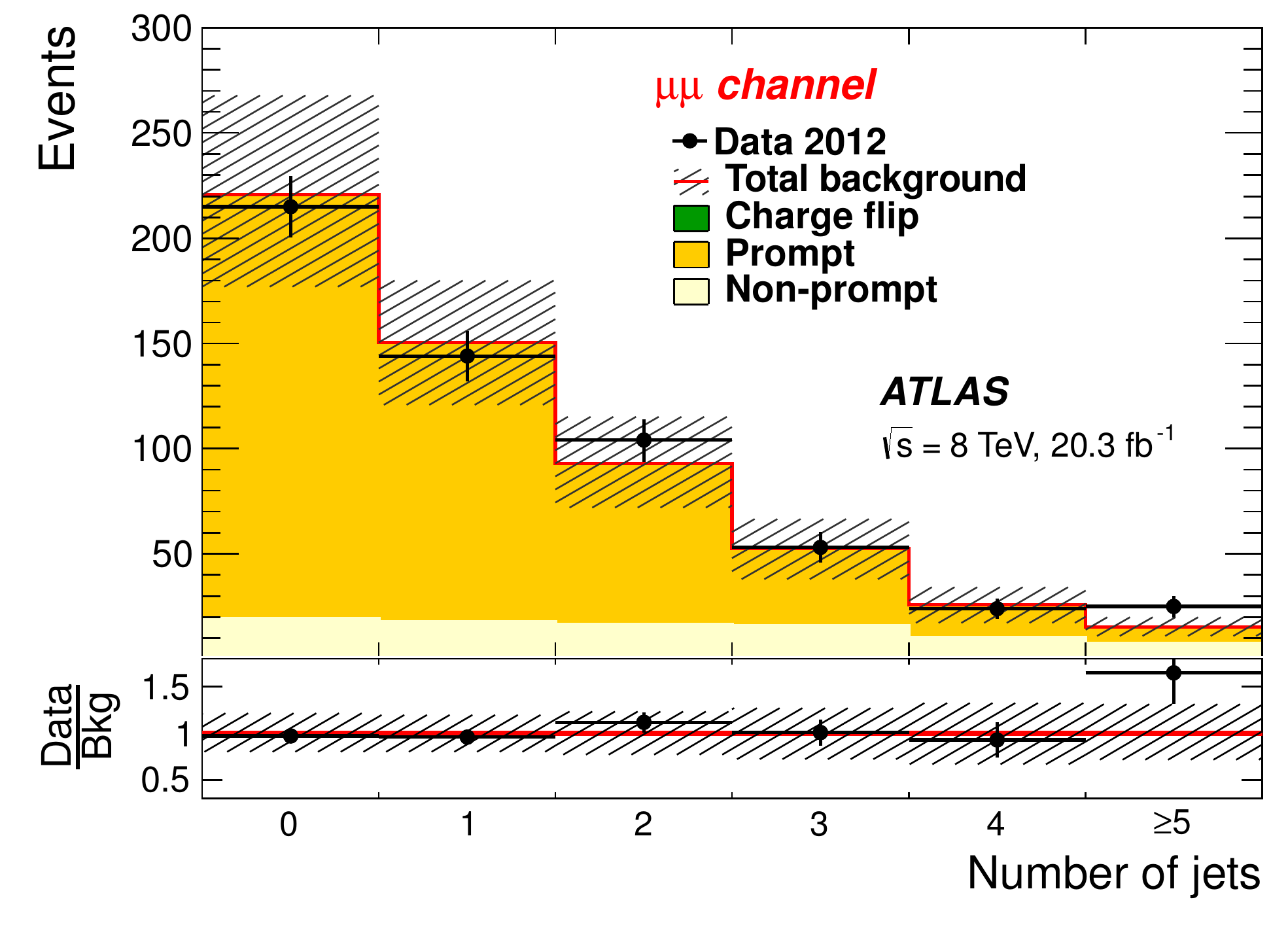}}
  \caption{The distribution of the transverse momentum $\pt$ of the leading jet (top) and
	the distribution of the number of jets (bottom) for the $ee$ (left) and $\mu\mu$ (right) channels in a validation region consisting of events with exactly two same-sign leptons.
The shaded bands indicate the total uncertainty, including all contributions described in section \ref{sys}, on the total expected background and
  the lower plots show the ratio of data to the total expected background.}
 \label{fig:0pjetcr}
\end{figure}

\clearpage
\section{Systematic uncertainties}
\label{sys}

The background estimates and signal efficiencies are subject to several systematic uncertainties.
The relative size of the uncertainties on the total background estimates in the mTISM and LRSM signal regions are detailed in tables \ref{Wmass60to100err} and \ref{LRSMerr} respectively.
Since the tables show the size of each uncertainty relative to the total background estimate, the impact of each uncertainty depends on the background composition in the different channels (see section~\ref{sec:results} and tables~\ref{tab:mTISMyields} and~\ref{tab:LRSMyields}).

\renewcommand{\arraystretch}{1.2}
\begin{table}[p]
\centering 
\small
\begin{tabular}[h]{l | c | c } 
 & \myalign{c|}{$ee$} & \myalign{c}{$\mu\mu$} \\ \hline 
Non-prompt & $\pm$ 7 & $\pm$ 14\\ 
Charge-flip & $\pm$ 7 & -- \\ 
Prompt normalisation & $\pm$ 2 & $\pm$ 10 \\ 
MC statistics & $\pm$ 14 & $\pm$ 8\\ 
Jet energy scale & +8/$-$22 & +7/$-$12\\ 
\met & +2/$-$3 & +6/$-$3 \\ 
Jet energy resolution & $\pm$ 4 & $\pm$ 3\\ 
Jet vertex fraction & +2/$-$6 & +4/$-$5\\ 
Lepton uncertainties & $\pm$ 2 & +2/$-$3\\ 
Luminosity & $\pm$ 2 & $\pm$ 1\\
\hline Total & +20/$-$29 & +21/$-$24 \\ \hline 
 \end{tabular} 
\caption{A breakdown of the relative uncertainty on the total background (given in \%) in the mTISM signal region. The various sources of systematic uncertainty are described in Section \ref{sys}.} 
\label{Wmass60to100err} 
\end{table} 

\renewcommand{\arraystretch}{1.2}
\begin{table}[p]
\centering 
\small
\begin{tabular}[h]{l | c | c | c | c } 
 & \multicolumn{2}{c|}{$ee$} & \multicolumn{2}{c}{$\mu\mu$} \\  
 & \WR & \ZR &\WR & \ZR  \\ \hline
Non-prompt & $\pm$ 12 & $\pm$ 15 & +10/$-$9 & +13/$-$11  \\ 
Charge-flip & +5/$-$4 & +4/$-$3 & -- & -- \\ 
Prompt normalisation & $\pm$ 5 & +5/$-$4 & $\pm$ 18 & $\pm$ 14  \\ 
MC statistics & $\pm$ 7 & $\pm$ 6 & $\pm$ 5 & $\pm$ 4  \\ 
Jet energy scale & +8/$-$9 & +5/$-$4 & $\pm$ 5 & +5/$-$4  \\ 
Jet energy resolution & $\pm$ 0.9 & $\pm$ 0.6 & $\pm$ 1.2 & $\pm$ 0.2  \\ 
Jet vertex fraction & $\pm$ 1 & $\pm$ 2 & +0.9/$-$0.1 & +2.4/$-$1.1  \\ 
Lepton uncertainties & +2/$-$1 & +2.7/$-$1 & $\pm$ 2 & $\pm$ 2  \\ 
Luminosity & $\pm$ 1 & +1.5/$-$1 & $\pm$ 1 & $\pm$ 1  \\ 
\hline Total & $\pm$ 18 & $\pm$ 18 & $\pm$ 21 & +20/$-$19  \\ \hline 
 \end{tabular} 
\caption{A breakdown of the relative uncertainty on the total background (given in \%) in the LRSM signal regions.  The various sources of systematic uncertainty are described in Section \ref{sys}. } 
\label{LRSMerr} 
\end{table}

\subsection{Background uncertainties}

The systematic uncertainty on the estimate of the non-prompt background is dominated by the uncertainties on the measurements of the rates for leptons
from fake and non-prompt sources
to satisfy the lepton identification criteria. This uncertainty is dominated by the effect of the choice of control region definition on the measured probabilities, as
well as the statistical uncertainty on determining the probabilities. 
The total uncertainty on the non-prompt background in the $ee$~$(\mu\mu)$~channel varies from $30-48\%$~$(41-45\%)$, depending on the signal region.

The charge-flip background is only relevant for the $ee$~channel (see section~\ref{sec:cf}) and its uncertainty is dominated by the statistical precision with which the charge-flip 
rate is determined from the available data. There is, additionally, a non-negligible contribution to the total uncertainty
due to the modelling of the subtraction of the non-prompt background, 
although this is correlated with the uncertainty on the non-prompt background. Since the charge-flip background uses MC simulation,
the systematic uncertainties discussed in section~\ref{sec:systMC} also apply to the charge-flip background.
The total uncertainty on the charge-flip background varies from $18-46\%$, depending on the signal region.

The uncertainty on the normalisation of the backgrounds originating from $WZ$~and $ZZ$~processes
is derived from the diboson control region described in section~\ref{sec-prompt}. This uncertainty is taken to be either the difference between the data
and the prediction or the statistical uncertainty from the limited data statistics, whichever is largest. The uncertainty is applied
as a function of the number of jets, leading to uncertainties of $10-14\%$, depending on the signal region. The uncertainty on the cross-section of
other background processes from MC estimates is taken from the uncertainties on the theoretical cross-sections.  The combined effect
of the diboson normalisation and theoretical cross-section uncertainties is labelled as `Prompt normalisation' in tables \ref{Wmass60to100err}~and \ref{LRSMerr}.

Another source of uncertainty 
is due to the MC statistical uncertainty. In the mTISM signal region this is particularly large in the $ee$ channel due to 
the small number of events in the $Z \rarr ee$ MC sample in this
region. 

\subsection{Uncertainties on MC simulation}
\label{sec:systMC}

The following uncertainties are applied to all MC-derived predictions. In addition to affecting the prompt and charge-flip background estimates, these
uncertainties also apply to the signal simulation.

The systematic uncertainty on the jet energy scale has an important effect on both the signal and background processes, as the signal-region event selection
includes requirements on quantities that are reconstructed from jet kinematics \cite{jes_unc}. The uncertainty on the jet energy scale is notably
asymmetric in the mTISM signal region (table~\ref{Wmass60to100err}) compared to the LRSM signal region (table~\ref{LRSMerr}) due to the dependency
on the leading dijet mass in the mTISM signal region definition described in section~\ref{sec:mTISM_sigsel}. The uncertainty due to the JES on the 
sum of the prompt and charge-flip backgrounds varies from $6-28\%$, depending on the signal region.
The modelling of the missing transverse momentum (`$\met$')~\cite{met} is included as a systematic uncertainty in the mTISM signal region (table~\ref{Wmass60to100err})
as there is a dependence on this quantity in the signal region definition described in section~\ref{sec:mTISM_sigsel}. 
Other smaller systematic uncertainties include the uncertainty on the the jet energy resolution~\cite{jersys}, the jet vertex fraction requirement, uncertainties
on lepton identification efficiencies, energy / momentum scales and resolutions (`Lepton uncertainties')~\cite{muonsys,Egamma_ID} and the uncertainty on the luminosity measurement (`Luminosity')~\cite{lumisys}.

\subsection{Signal-specific modelling uncertainties}

In addition to the uncertainties associated with the MC simulation of background processes, there are systematic modelling uncertainties
associated with the signal MC samples.
An uncertainty is considered for the signal MC simulation 
to reflect the choice of parton shower model. The nominal parton shower model that is used for all 
signal MC samples is {\sc Pythia 8.165}. 
The total number of events in the signal region when the signal MC generator is interfaced to {\sc Pythia} is compared to the number of events when the generator is interfaced to {\sc Herwig 6.520}.
The variation in the signal efficiency is measured to be approximately~$5\%$. 
The uncertainty due to the parton distribution functions on the signal acceptance is found to be approximately $5\%$~for the mTISM signal samples
and approximately $7\%$~for the LRSM signal samples.

A systematic uncertainty is also considered to cover the effect of using the fast detector simulation described in section~\ref{sec-MC}.
Two versions of the MC signal, one with a full detector simulation and the other with the standard fast detector simulation, are compared in each lepton channel, for a single signal mass point.
The difference
in the efficiency to select signal events, approximately $4\%$, is assigned as an uncertainty.

\section{Results}
\label{sec:results}

The numbers of events measured in data are compared to the expected numbers of background events in the signal regions, with the intention of interpreting an excess of events in data in terms of a heavy Majorana neutrino in the mTISM or LRSM models. 

\subsection{Results in the mTISM signal region}

{The observed and predicted distributions of the invariant mass of the two highest-$\pt$ jets ($m_{jj}$) in events satisfying the mTISM selection criteria, excluding the criteria on 
 $m_{jj}$ (60~GeV$ < m_{jj} < 100$~GeV)
are presented in figure \ref{fig:mTISMSR}. The shapes of the distributions show good agreement between data and expectation. The numbers of expected and observed events in the mTISM signal region (indicated by arrows in figure \ref{fig:mTISMSR}) are shown in table~\ref{tab:mTISMyields}. There is no excess of events relative to the expectation. 
The observed yields in the data are used
to set $95\%$~CL upper limits on the production cross-section times branching ratio, $\sigma \times \mathrm{Br}(pp \to \ell^{\pm} N \to \ell^{\pm}\ell^{\pm} q\bar{q}')$,
of heavy neutrinos to electrons or muons, using the profile-likelihood test statistic~\cite{s10052-011-1554-0,erratum} and the
$CL_s$~method~\cite{0954-3899-28-10-313}.
The systematic uncertainties are included in the test statistic as nuisance parameters. Each systematic uncertainty is assumed to be uncorrelated with all other systematic uncertainties.
The limits are shown as a function of $\mN$ in figure~\ref{fig:mTISMlimits} and are translated into limits on the mixing parameter, $|V_{\ell N}|^2$, between the heavy neutrino
and the SM neutrino, separately for the $ee$ and $\mu\mu$ channels. The extraction of the limits on the mixing parameters
uses the leading-order cross-section for the signal process and no uncertainties are included on the signal cross-section.
The extraction furthermore assumes that only the lightest of the heavy neutrinos contributes to 
the cross-section and that the masses of the other heavy neutrino species are sufficiently high that the effect of interference is negligible. 
}

\begin{figure}[htbp]
\centering
\subfigure[]{\includegraphics[width=0.48\textwidth]{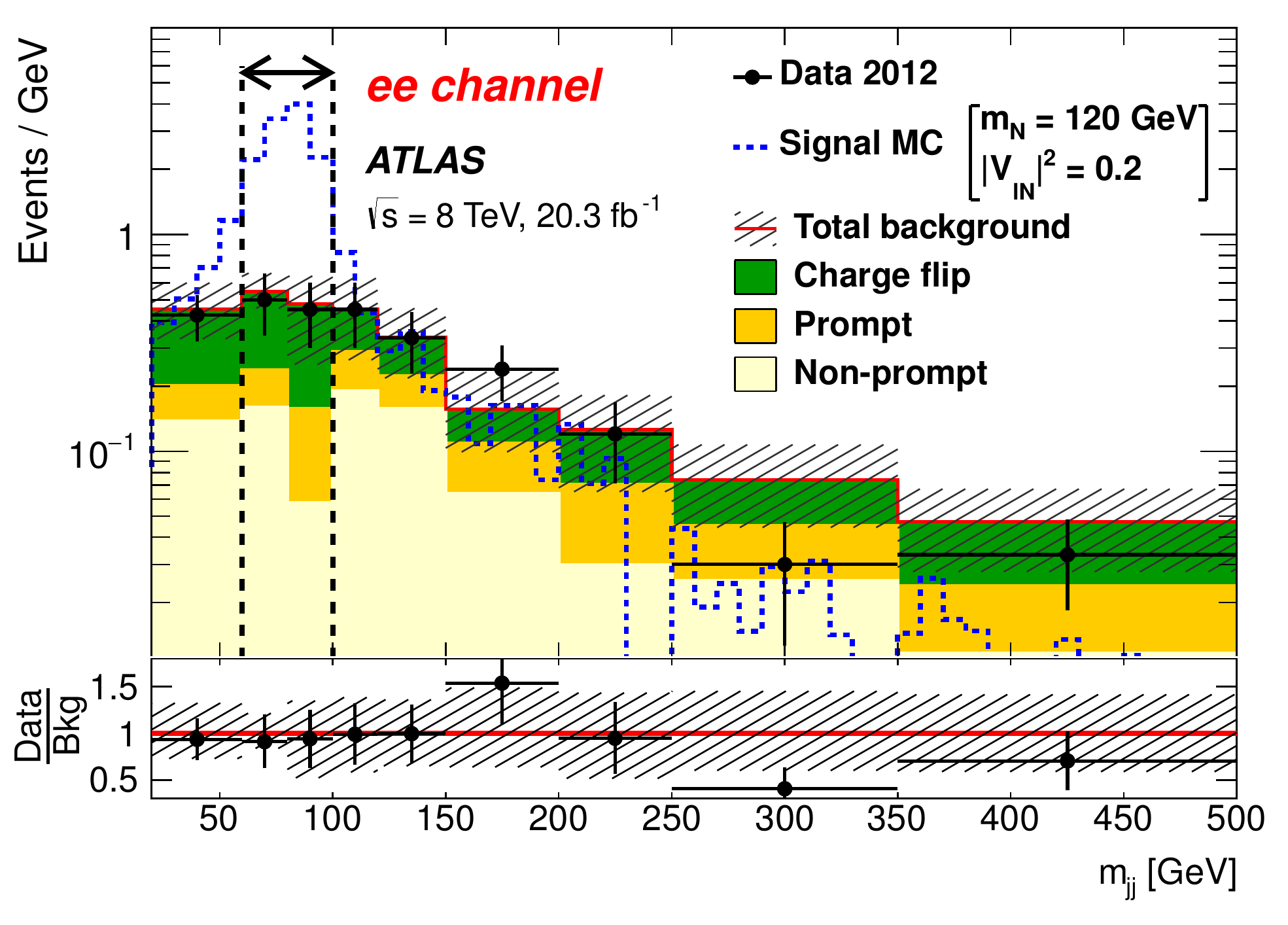}}
\subfigure[]{\includegraphics[width=0.48\textwidth]{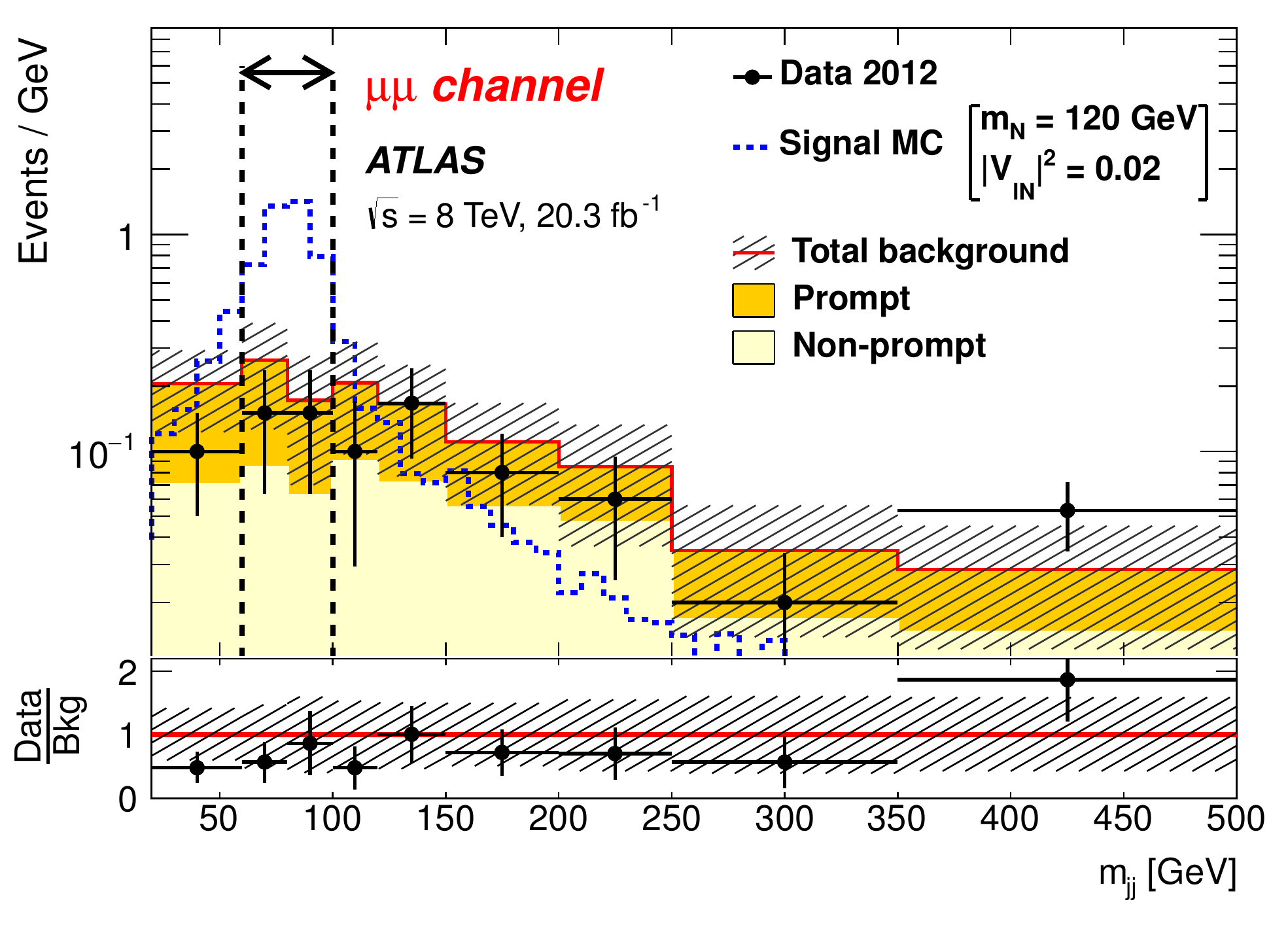}}
  \caption{Invariant mass of the two highest-$\pt$ jets ($m_{jj}$) in events satisfying the mTISM signal region criteria (excluding the $m_{jj}$ criteria) for (a) $ee$ and (b) $\mu\mu$ events. Events satisfying all selection criteria are in the region indicated by the arrows. The expected mTISM signal distribution for $m_N = 120 \GeV$ is shown by the dashed (blue) histogram. The values of the mixing parameter $|V_{\ell N}|^2$ are chosen such that the signal distribution is clearly visible, $|V_{e N}|^2=0.2$~for (a) and $|V_{\mu N}|^2=0.02$~for (b). The shaded bands indicate the total uncertainty, including all contributions described in section \ref{sys}, on the total expected background and
  the lower plots show the ratio of data to the total expected background.}
  \label{fig:mTISMSR}
\end{figure}

\begin{table}[htbp]
\renewcommand{\arraystretch}{1.2} 
\centering
\small
\begin{tabular}{ l | c | c  } 
                & $ee$ & $\mu\mu$ \\
 \hline
Prompt	 & $3.5^{+0.9}_{-1.7}$	 & $5.8^{+1.3}_{-1.7}$	\\
Charge-flip	 & $13^{+3}_{-6}$	 & $< 0.02$	\\
Non-prompt	 & $4.3 \pm 1.8$	 & $2.9 \pm 1.3$	\\
 \hline
Total background	 & $21^{+4}_{-6}$	 & $8.7 \pm 2.0$	\\
  \hline
  Data & $19$ & $6$ \\
 \hline
  Signal ($m_N = 120 \GeV$) & $18$  & $6.6$ \\
  Signal ($m_N = 240 \GeV$) & $30$ & $5.3$  \\
  \hline
\end{tabular} 
\caption{Total event yields measured in data and predicted for signal and background processes in the mTISM signal region. The uncertainties shown on the various backgrounds correspond to the total uncertainty. The expected number of mTISM signal events are calculated for $m_N = 120 \GeV$ (with $|V_{\ell N}|^2$ equal to 0.03 and 0.003 in the $ee$ and $\mu\mu$ channels respectively) and $m_N = 240 \GeV$ (with $|V_{\ell N}|^2$ equal to 0.2 and 0.02 in the $ee$ and $\mu\mu$ channels respectively). The
values of the mixing parameters $|V_{\ell N}|^2$ are chosen to be close to the expected limit shown in figure~\ref{fig:mTISMlimits}.
For backgrounds that have zero expected events, the 95\% CL upper limit is shown.}
\label{tab:mTISMyields}
\end{table}

\begin{figure*}[hbtp]
\centering
\subfigure[]{\includegraphics[width=0.48\textwidth]{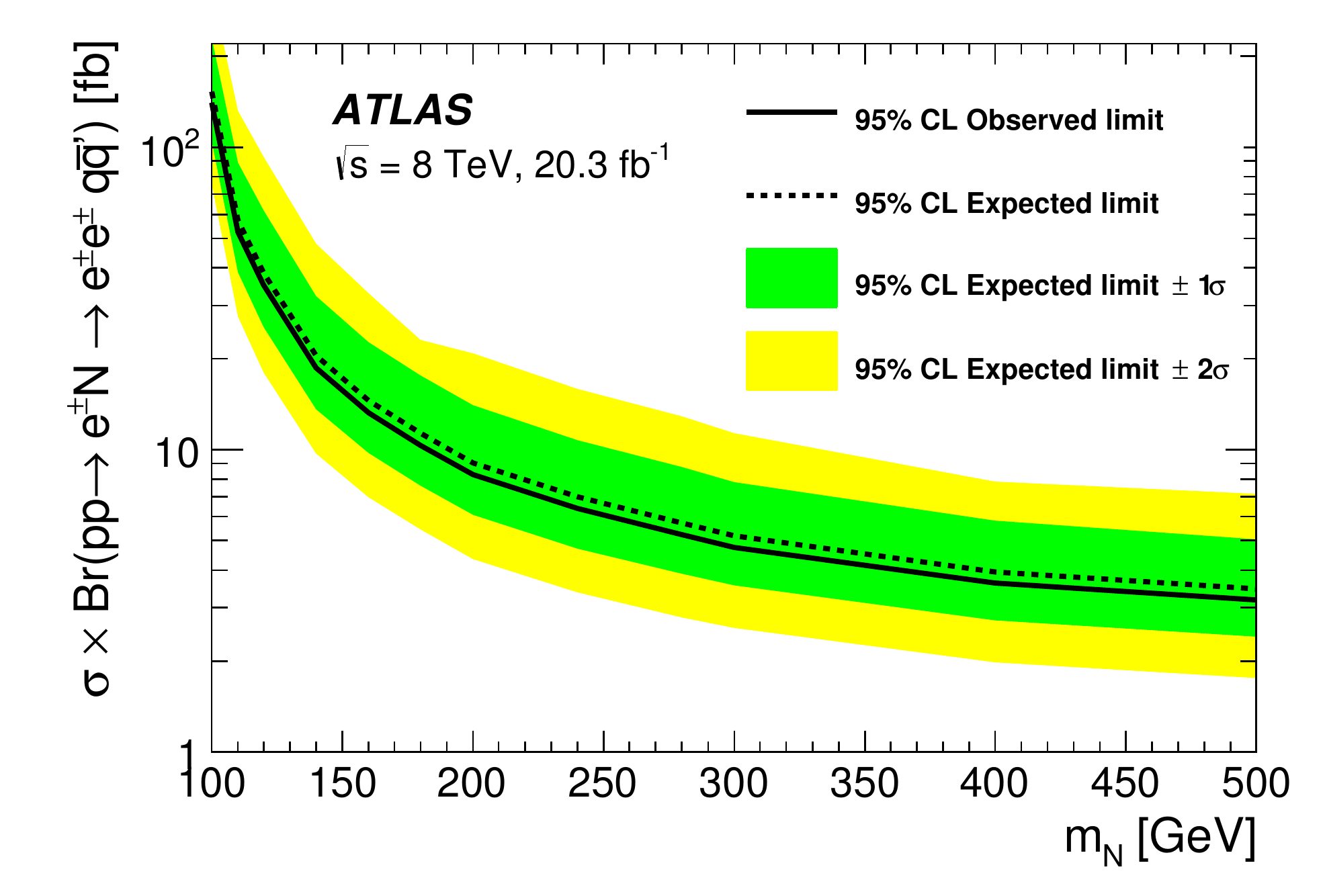}}
\subfigure[]{\includegraphics[width=0.48\textwidth]{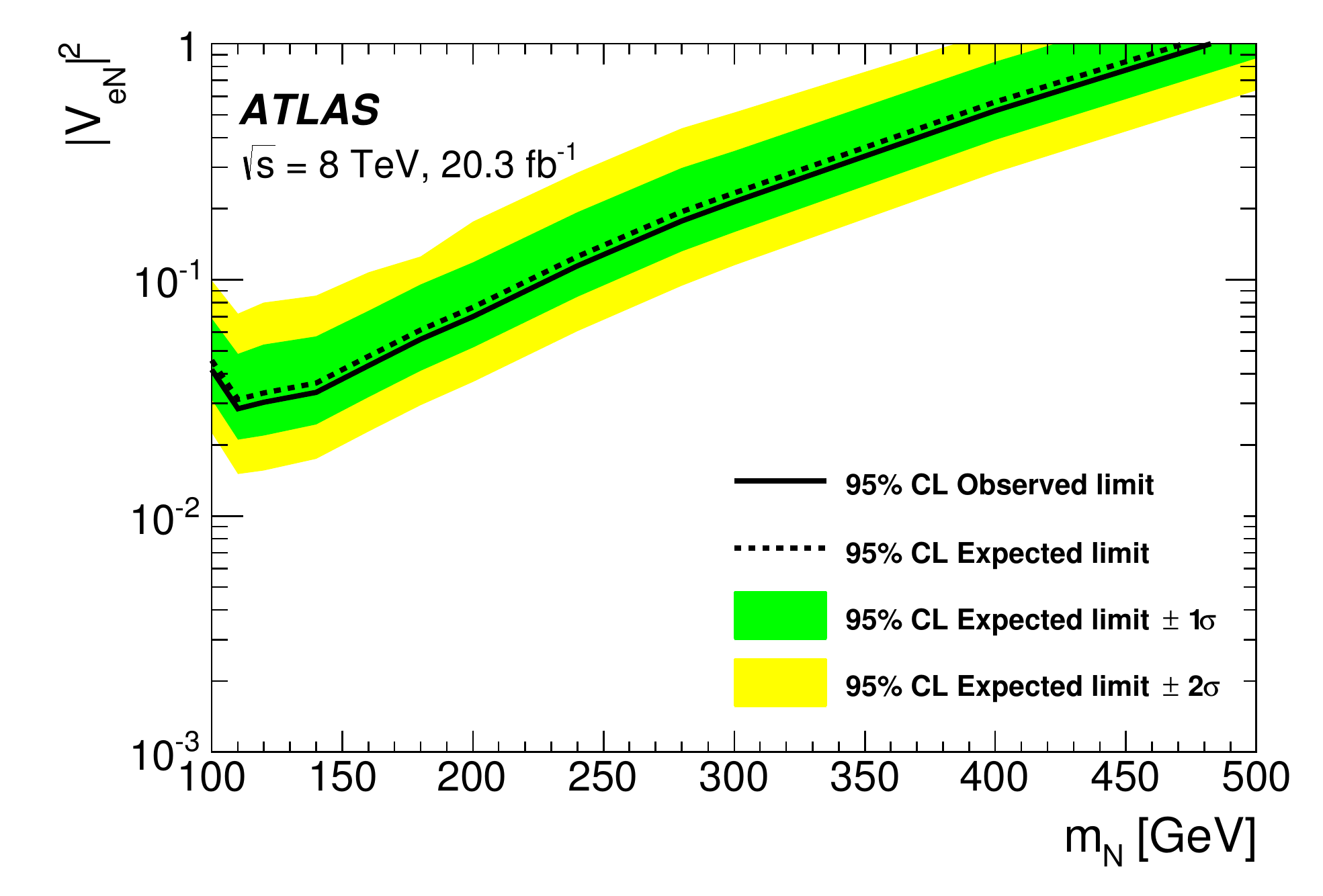}}
\subfigure[]{\includegraphics[width=0.48\textwidth]{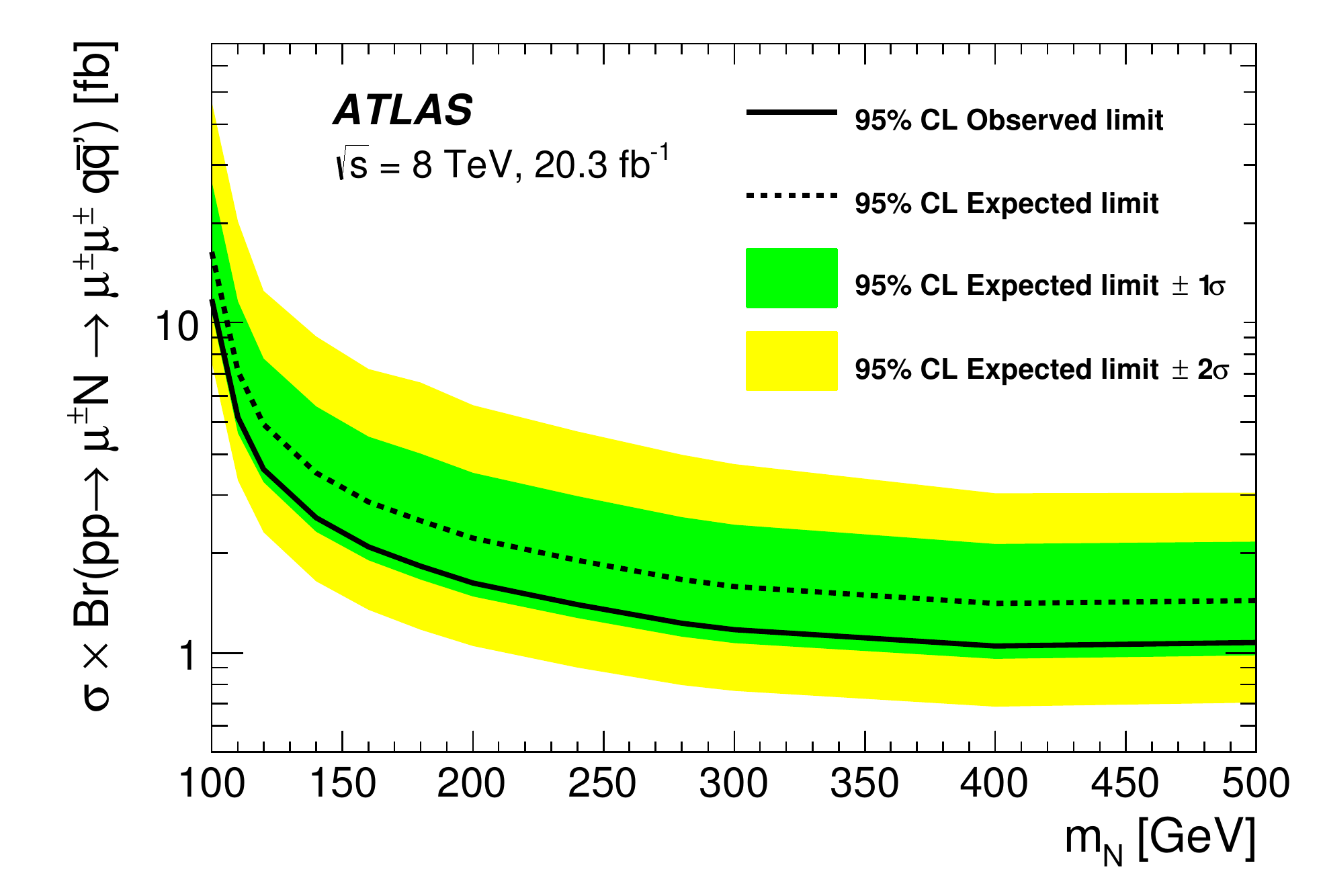}}
\subfigure[]{\includegraphics[width=0.48\textwidth]{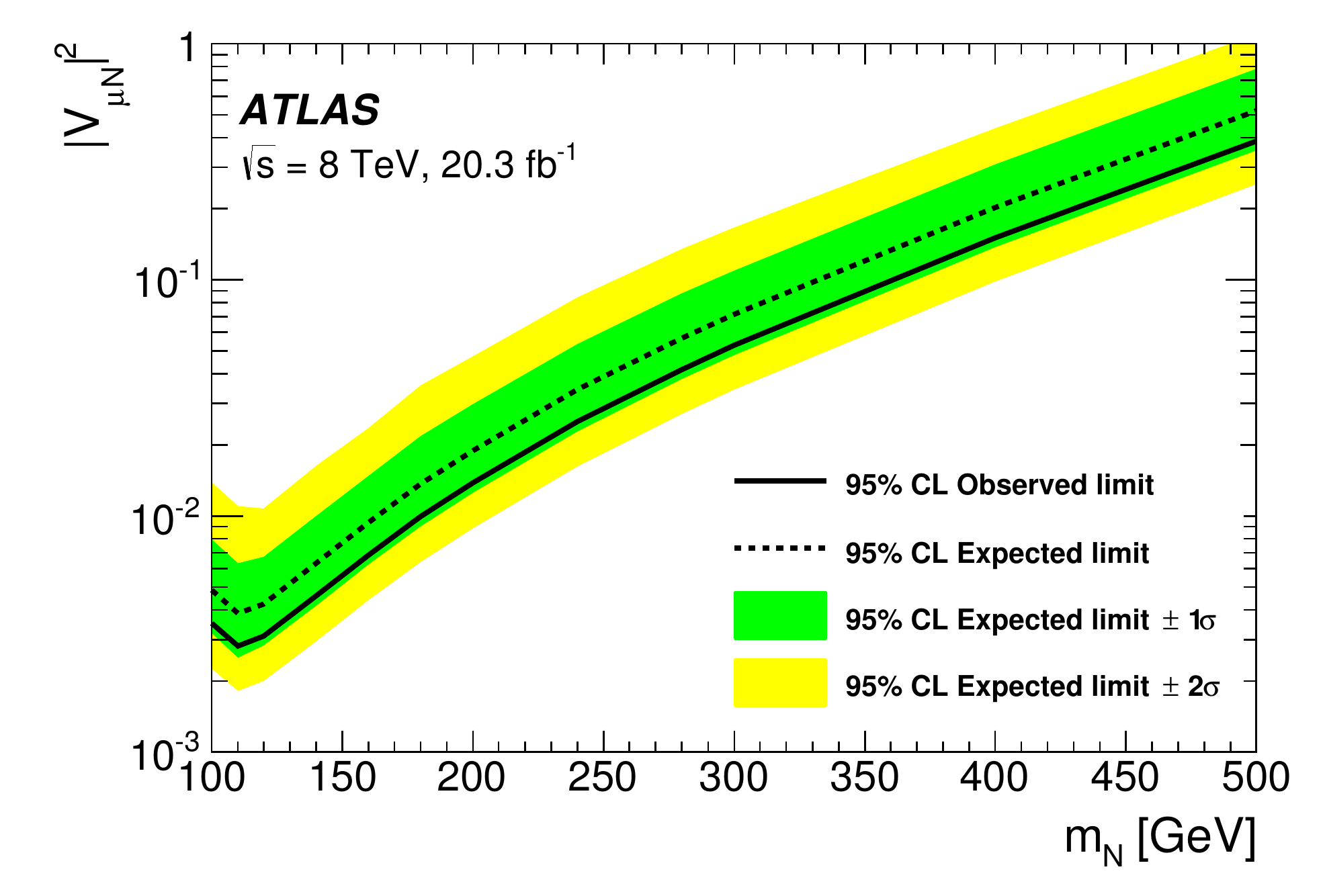}}
\caption{Observed and expected $95\%$~confidence level upper limits on the cross-section times branching ratio for the production of mTISM heavy Majorana neutrinos as a function of the heavy neutrino mass for (a) the $ee$~channel and (c) the $\mu\mu$~channel. The limits on the mixing between the heavy Majorana neutrinos and the SM neutrinos are shown in (b) and (d). Values larger than the solid black line are excluded by this analysis.}
  \label{fig:mTISMlimits}
\end{figure*}

\subsection{Results in the LRSM signal region}

The observed and expected numbers of events for the LRSM signal regions are shown in table~\ref{tab:LRSMyields}. There are no excesses observed above the expected numbers of background events. 

The LRSM signal is expected to produce a peak in the invariant mass of the decay
products of the heavy gauge boson. This would be observed in the invariant mass 
 distribution $m_{\ell \ell j(j)}$ ($m_{\ell \ell jj(jj)}$) in the \WR~(\ZR)
signal regions, as described in section \ref{evsel}. The observed and
predicted distributions are shown in figures~\ref{fig:LRSMWr} and~\ref{fig:LRSMZr}.
Binned likelihood fits are performed to the invariant mass distributions and the profile-likelihood test statistic is used to assess the compatibility of the data with the background-only and signal-plus-background hypotheses. 
No significant excess is observed in the data compared to the background expectation and $95\%$~CL upper limits on the cross-section of the production
of heavy gauge bosons decaying to heavy neutrinos within the LRSM are set using the $CL_s$~method.
The expected and observed cross-section exclusion limits as a function of the masses of the heavy gauge bosons and
heavy neutrino are shown for example mass points
for both channels, $ee$ and $\mu\mu$, in table~\ref{tab:LRSMlimitsSSSel}. The full cross-section limits for all analysed mass points are available in HepData.\footnote{http://hepdata.cedar.ac.uk/}
Exclusion contours are also set in the $\{ \mVR$,\ $\mN \}$ plane, and are shown in figure~\ref{fig:LRSM_limit_Majorana_SS}.
The contours are found by comparing the cross-section limits to the leading-order cross-sections for the signal processes and no uncertainties are included on the signal cross-sections.
For this interpretation there is assumed to be no mixing between lepton flavours and three cases are investigated: the first two cases
assume a single heavy neutrino being kinematically accessible and being of either electron or muon flavour, which leads to events expected in one of the
$ee$~and $\mu\mu$~channels, respectively. The limits are stronger in the $\mu\mu$~channel at low heavy neutrino mass, due to the higher signal acceptance discussed in Section~\ref{sec:LRSM_sigsel}. As the efficiency of the LRSM signal is low in the ee channel for $\mVR \gg \mN$, the limit in this region is weaker.
Finally the case where two degenerate neutrinos are present is investigated, leading to events expected in both the $ee$~and $\mu\mu$~channels. The expected limits for this scenario
are slightly stronger than either of the individual channel scenarios. 
The sharp changes in the expected and observed limits visible in figure~\ref{fig:LRSM_limit_Majorana_SS} originate from the interpolation between the limited number of
mass points for which the limits are extracted.

\begin{table}[htb]
\renewcommand{\arraystretch}{1.2} 
\centering
\small
\begin{tabular}{ l | c | c | c | c   } 
                & \multicolumn{2}{c|}{$ee$} & \multicolumn{2}{c}{$\mu\mu$} \\
                & \WR & \ZR & \WR & \ZR \\
 \hline
 Prompt	 & $26 \pm 5$	 & $34 \pm 6$	 & $33 \pm 8$	 & $42 \pm 10$	 \\
Charge-flip	 & $44 \pm 11$	 & $44^{+10}_{-8}$	 & $<0.03$	 & $<0.03$	\\
Non-prompt	 & $23 \pm 11$	 & $33^{+11}_{-10}$	 & $9.8^{+5}_{-4}$	 & $17^{+8}_{-7}$	 \\ \hline
Total background	 & $93 \pm 16$	 & $111^{+16}_{-14}$	 & $43 \pm 9$	 & $60^{+13}_{-12}$	 \\
  \hline
  Data & $94$ & $106$ & $44$ & $55$ \\
 \hline
  Signal & $2.4$ & $5.2$ & $2.9$ & $7.7$ \\
    \hline
\end{tabular} 
\caption{Total event yields measured in data and predicted for signal and background processes in the LRSM signal regions. The uncertainties shown on the various backgrounds correspond to the total uncertainty.
The number of LRSM signal events are calculated for $\{m_{W_{\text{R}}}$; $m_{N}\}$ = \{2600; 1950\}~GeV and $\{m_{\ZR}$; $m_{N}\}$ = \{2200; 550\}~GeV.
For backgrounds that have zero expected events, the 95\% CL upper limit is shown.}
\label{tab:LRSMyields}
\end{table}

\begin{figure}[htbp]
\centering
\subfigure[]{\includegraphics[width=0.48\textwidth]{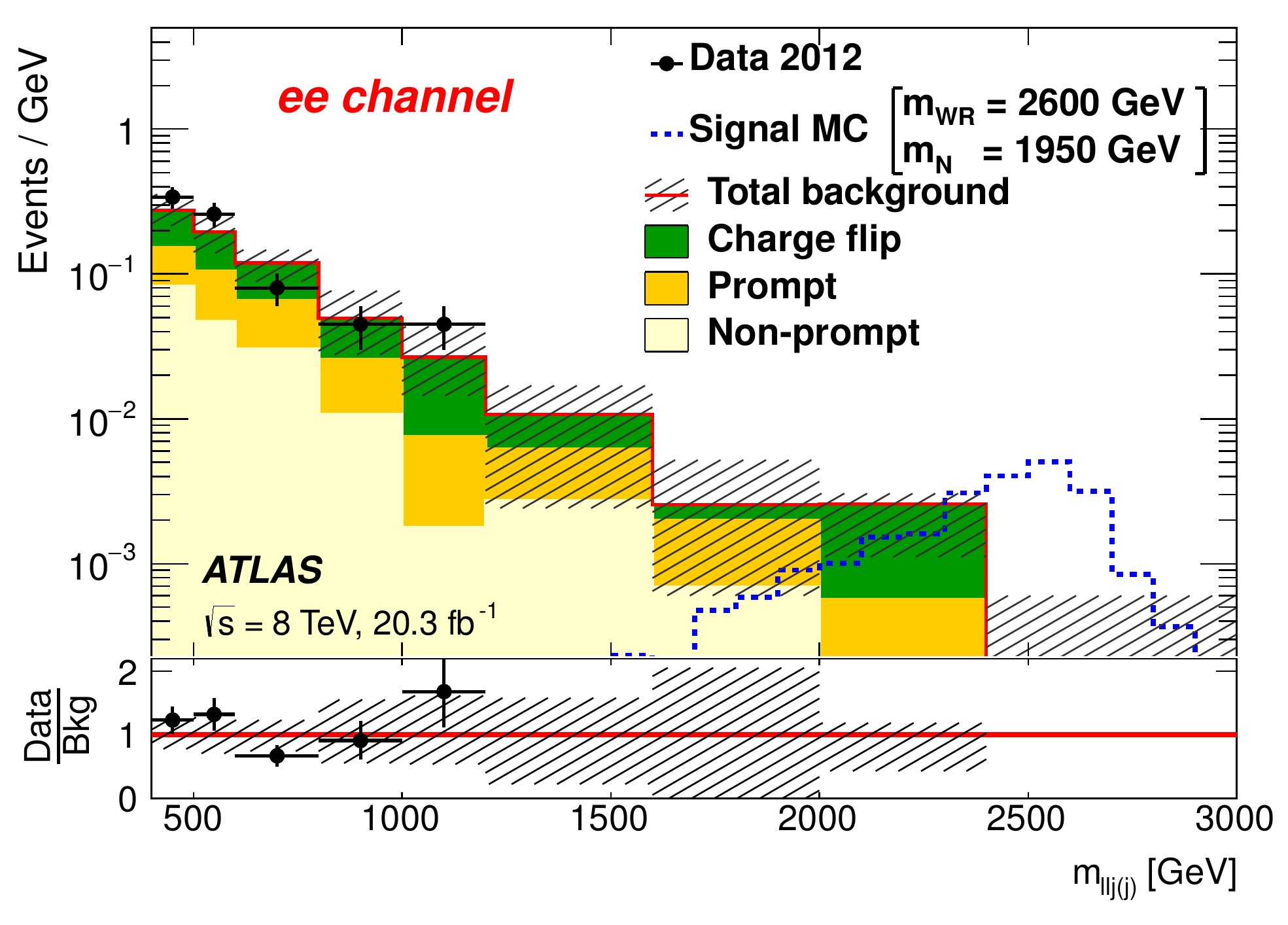}}
\subfigure[]{\includegraphics[width=0.48\textwidth]{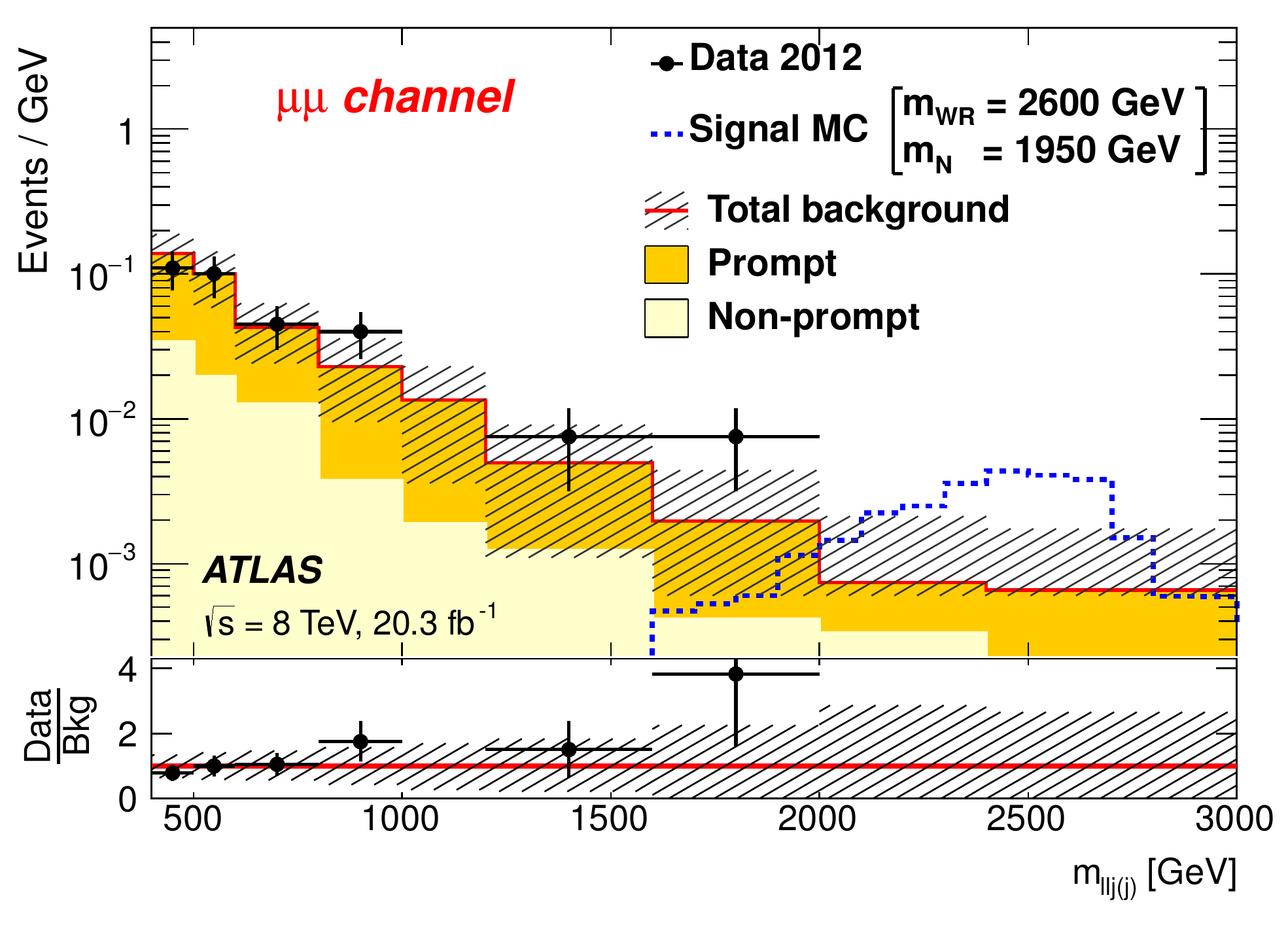}}
  \caption{Invariant mass of two leading leptons and up to two leading
  jets after applying additional $W_{\text{R}}$ selection criteria (two same-sign leptons, at least one jet, $m_{\ell\ell} > 110 \GeV$ and $m_{\ell \ell j(j)} > 400 \GeV$), for the $ee$-channel (a) and $\mu\mu$-channel (b). A finely binned 
  LRSM signal sample is represented by the dashed (blue) histogram
  corresponding to $\mWR = 2600 \GeV$ and $\mN = 1950 \GeV$. The shaded bands indicate the total uncertainty, including all contributions described in section \ref{sys}, on the total expected background and
  the lower plots show the ratio of data to the total expected background.}
  \label{fig:LRSMWr}
\end{figure}

\begin{figure}[htbp]
\centering
\subfigure[]{\includegraphics[width=0.48\textwidth]{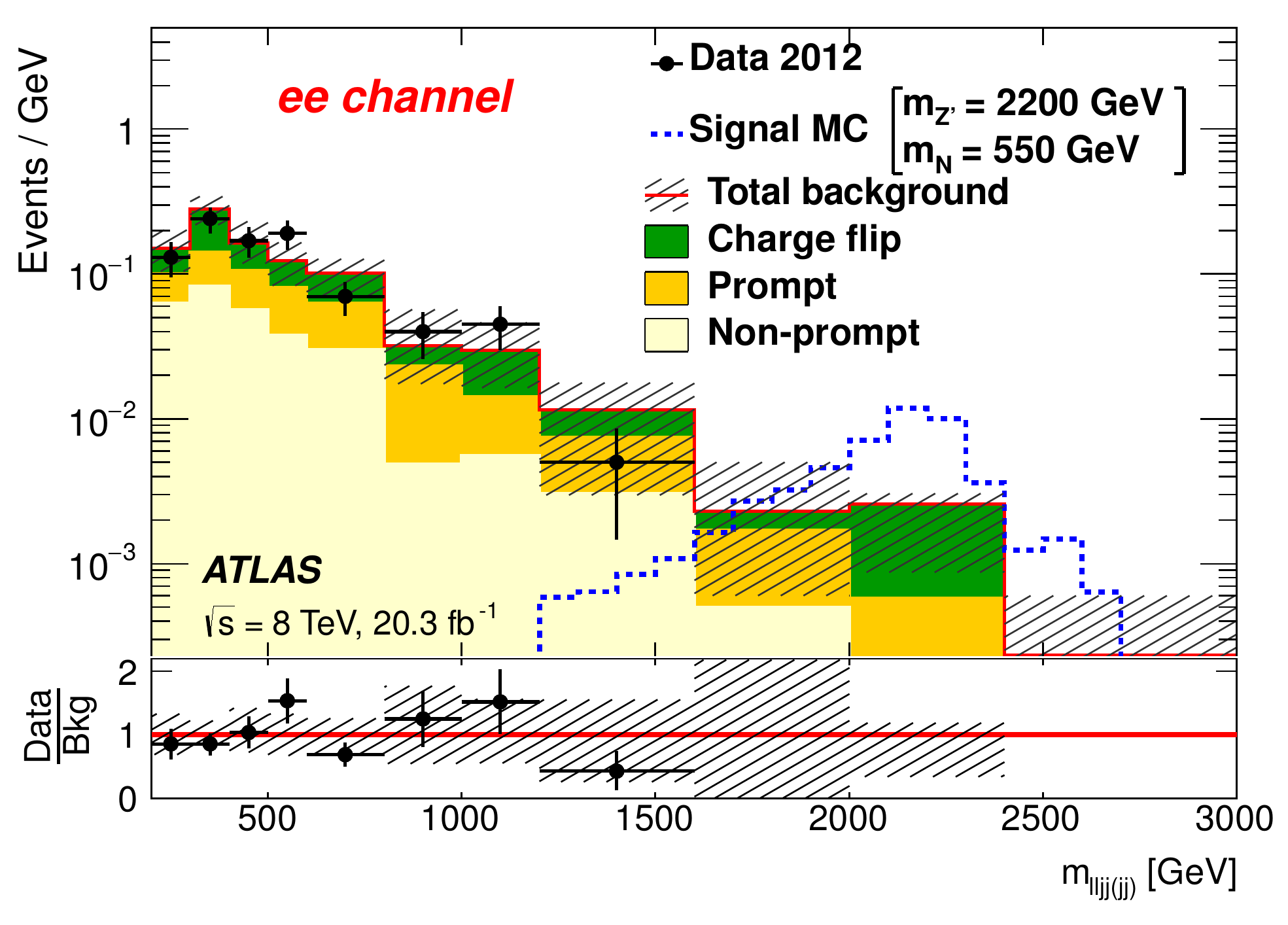}}
\subfigure[]{\includegraphics[width=0.48\textwidth]{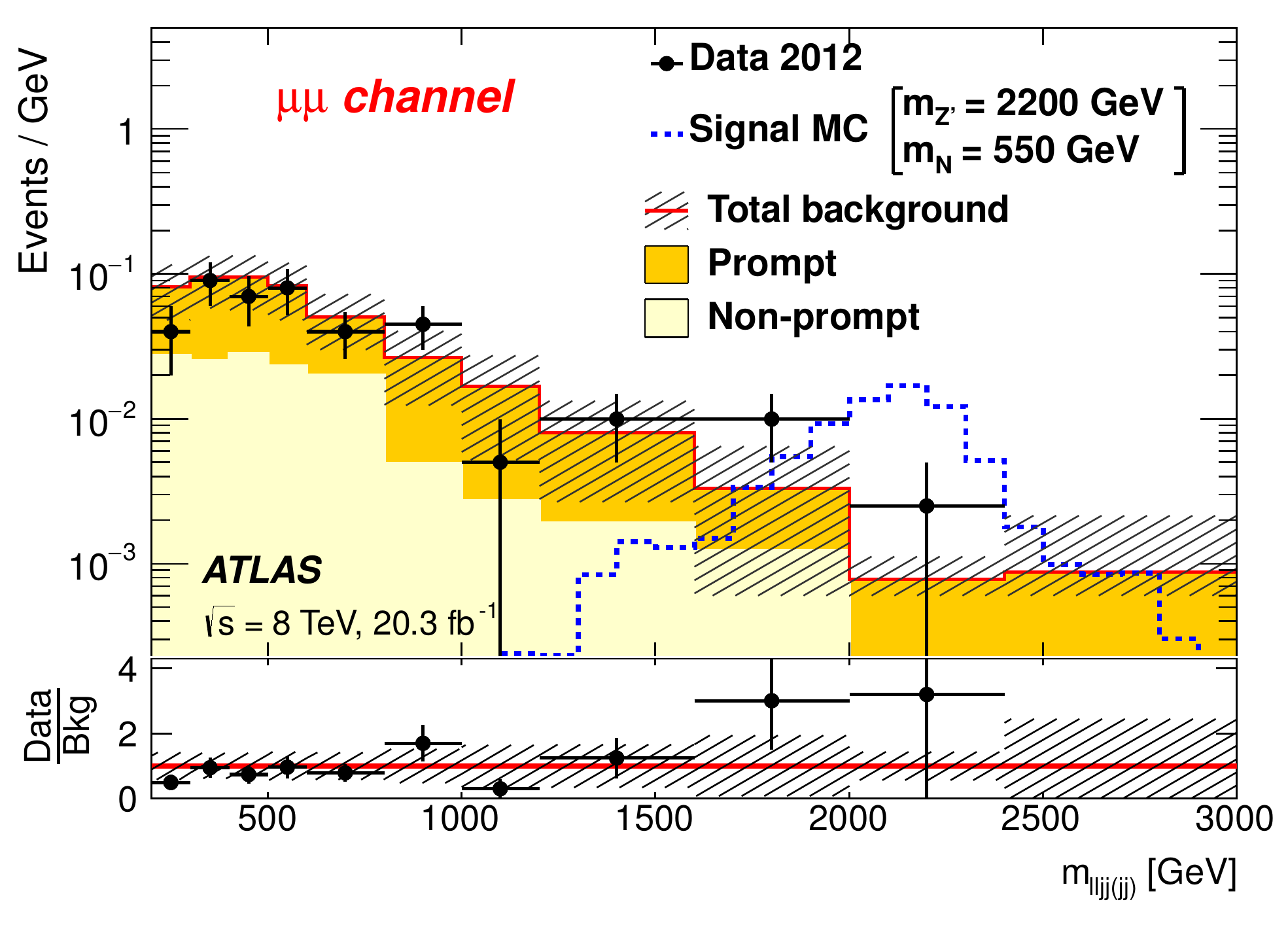}}
  \caption{Invariant mass of two leading leptons and up to four leading
  jets after applying additional $\ZR$ selection criteria (two same-sign leptons, at least two jets, $m_{\ell\ell} > 110 \GeV$ and $m_{\ell \ell jj(jj)} > 200 \GeV$), for the $ee$-channel (a) and $\mu\mu$-channel (b). A finely binned 
  LRSM signal sample is represented by the dashed (blue) histogram
  corresponding to $\mZR = 2200 \GeV$ and $\mN = 550 \GeV$. The shaded bands indicate the total uncertainty, including all contributions described in section \ref{sys}, on the total expected background and
  the lower plots show the ratio of data to the total expected background.}
  \label{fig:LRSMZr}
\end{figure}

\clearpage

\begin{table}[htb]
\begin{center}
\begin{tabular}{|c|c|c|c|c|c|c|}
\hline $m_{\VR}$ & $m_{N}$ & \multirow{2}{*}{Efficiency $[\%]$} & Expected & Expected & Expected  & Observed  \\ 
 $[\mathrm{GeV}]$ & $[\mathrm{GeV}]$ & & limit [fb] & limit $\pm 1\sigma$~[fb] & limit $\pm 2\sigma$~[fb] & limit [fb] \\ \hline\hline
\multicolumn{7}{|l|}{$\WR$~$ee$} \\ \hline
1400 & 50 & 0.80 & 100 & $130 / 70$ & $190 / 49$ & 65 \\
1400 & 1050 & 22 & 2.2 & $3.1 / 1.5$ & $3.8 / 1.1$ & 1.7 \\
2300 & 50 & 0.39 & 400 & $530 / 250$ & $780 / 160$ & 320 \\
2300 & 1725 & 23 & 0.87 & $1.1 / 0.64$ & $1.8 / 0.44$ & 0.56 \\
2900 & 50 & 1.2 & 140 & $210 / 98$ & $270 / 71$ & 160 \\
2900 & 2175 & 23 & 0.85 & $1.1 / 0.62$ & $1.7 / 0.43$ & 0.54 \\
\hline
\multicolumn{7}{|l|}{$\WR$~$\mu\mu$} \\ \hline
1400 & 50 & 19 & 1.8 & $2.8 / 1.2$ & $3.5 / 0.88$ & 2.1 \\
1400 & 1050 & 26 & 1.2 & $1.7 / 0.83$ & $2.2 / 0.59$ & 1.1 \\
2300 & 50 & 18 & 1.4 & $2.2 / 0.91$ & $3.2 / 0.63$ & 2.1 \\
2300 & 1725 & 26 & 0.85 & $1.3 / 0.55$ & $1.8 / 0.38$ & 1.3 \\
2900 & 50 & 15 & 2.2 & $3.5 / 1.4$ & $5.4 / 0.99$ & 3.4 \\
2900 & 2175 & 25 & 0.87 & $1.4 / 0.57$ & $1.8 / 0.39$ & 1.3 \\
\hline
\multicolumn{7}{|l|}{$\ZR$~$ee$} \\ \hline
1000 & 50 & 0.27 & 500 & $740 / 340$ & $920 / 240$ & 440 \\
1000 & 375 & 14 & 5.2 & $6.6 / 3.8$ & $8.3 / 2.8$ & 4.2 \\
2000 & 50 & 0.091 & 1100 & $1600 / 770$ & $2000 / 560$ & 1300 \\
2000 & 750 & 17 & 2.8 & $3.6 / 2.0$ & $4.8 / 1.5$ & 2.6 \\
2400 & 50 & 0.16 & 570 & $780 / 400$ & $970 / 290$ & 550 \\
2400 & 900 & 17 & 2.9 & $3.6 / 2.0$ & $4.8 / 1.5$ & 2.6 \\
\hline
\multicolumn{7}{|l|}{$\ZR$~$\mu\mu$} \\ \hline
1000 & 50 & 11 & 4.8 & $7.2 / 3.2$ & $10 / 2.2$ & 8.5 \\
1000 & 375 & 24 & 2.2 & $3.2 / 1.5$ & $4.6 / 1.1$ & 3.8 \\
2000 & 50 & 15 & 2.7 & $4.0 / 1.8$ & $5.8 / 1.3$ & 3.4 \\
2000 & 750 & 23 & 1.7 & $2.6 / 1.2$ & $3.3 / 0.83$ & 2.2 \\
2400 & 50 & 12 & 3.7 & $5.6 / 2.5$ & $7.0 / 1.8$ & 4.8 \\
2400 & 900 & 24 & 1.7 & $2.5 / 1.2$ & $3.3 / 0.82$ & 2.2 \\
\hline
\end{tabular}
\caption{Expected and observed 95\% CL limits for the production of LRSM $\WR$~or $\ZR$, with subsequent decay into a heavy Majorana neutrino in the $ee$~and $\mu\mu$~channels for several example mass points. Also shown is the event selection efficiency (including the detector acceptance). The full results for all analysed mass points are available in HepData.}
\label{tab:LRSMlimitsSSSel}
\end{center}
\end{table}

\begin{figure}[t]
  \centering
  \subfigure[]{\includegraphics[width=0.48\textwidth]{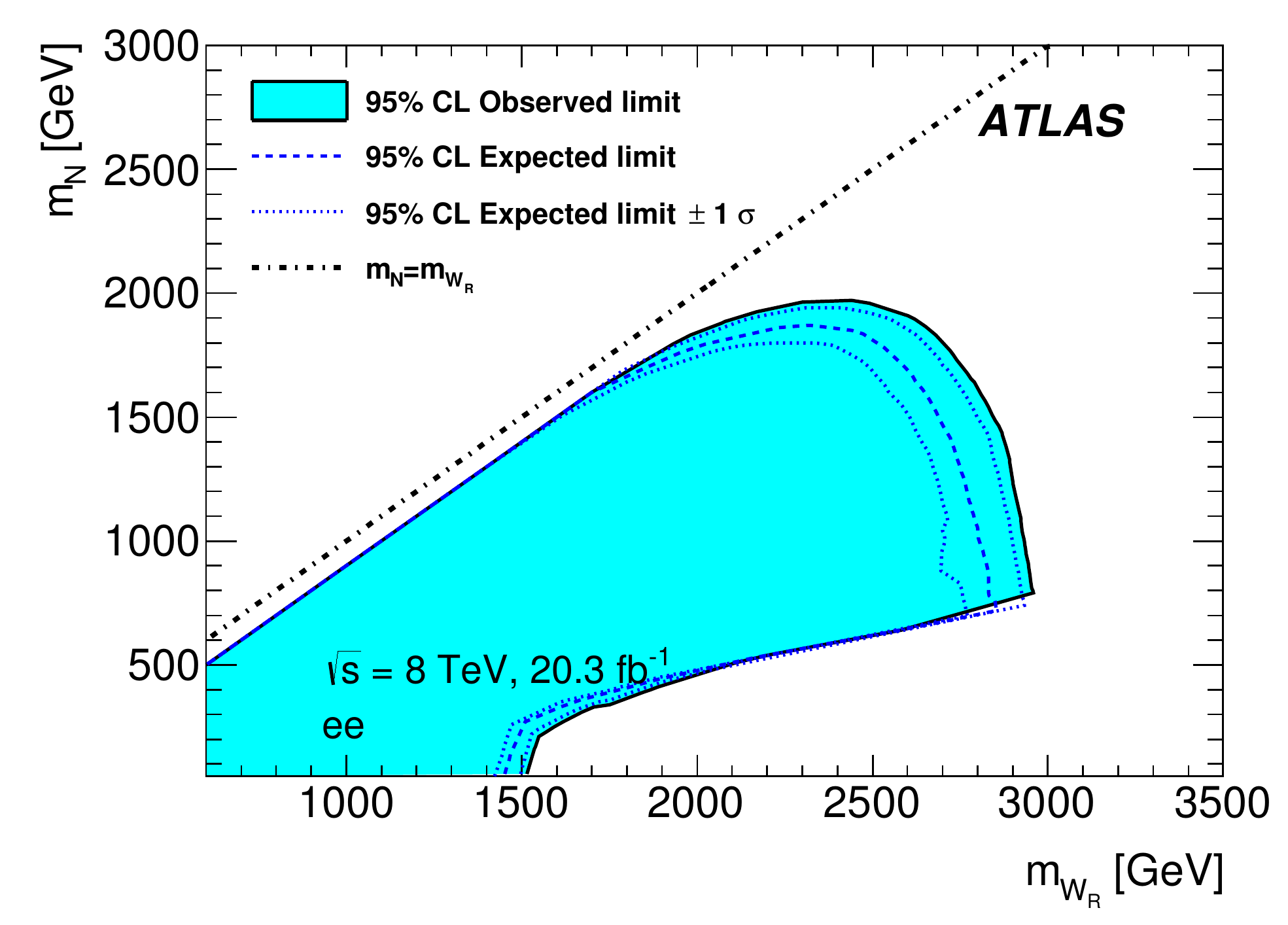}}
  \subfigure[]{\includegraphics[width=0.48\textwidth]{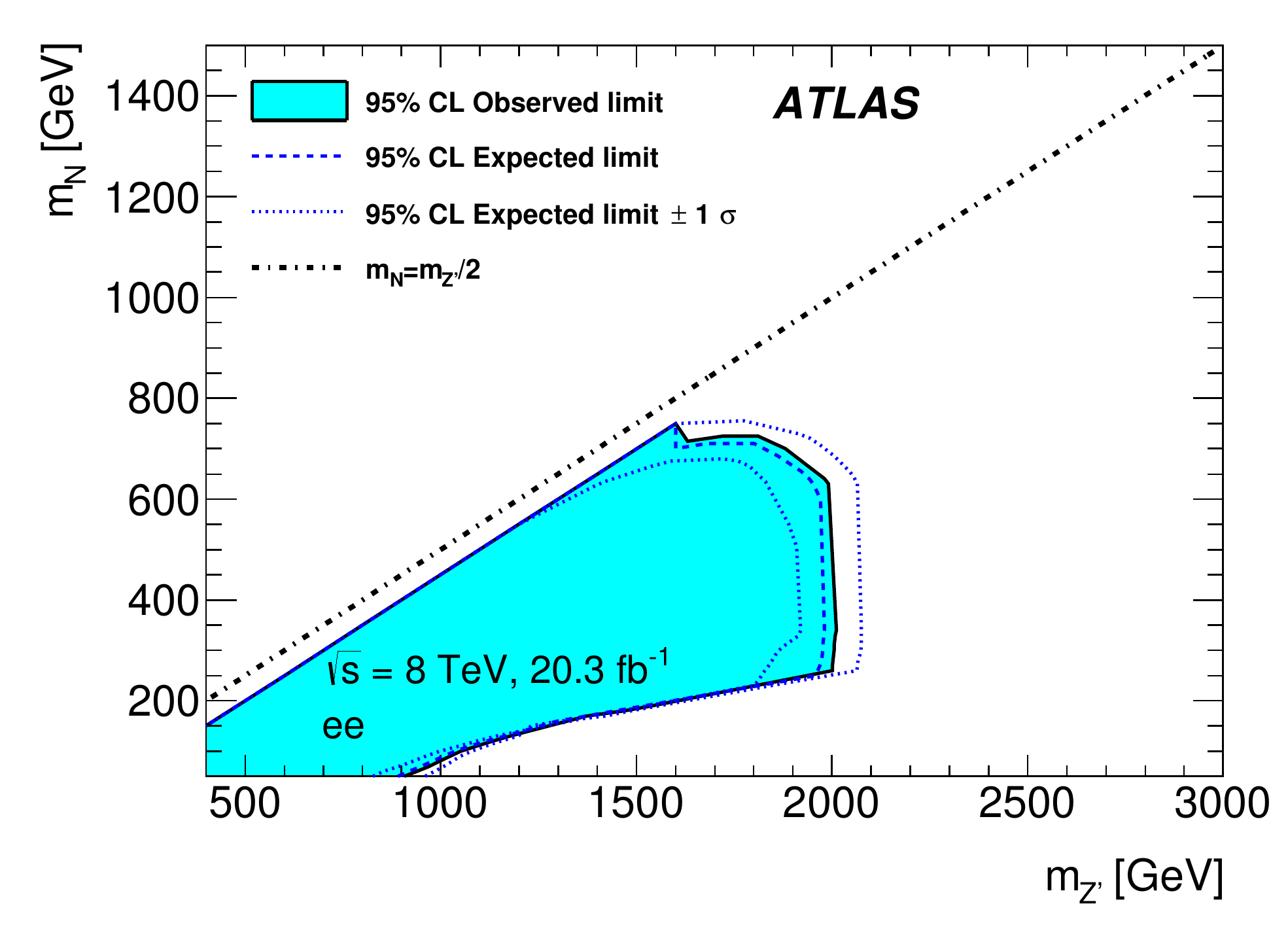}}
    \subfigure[]{\includegraphics[width=0.48\textwidth]{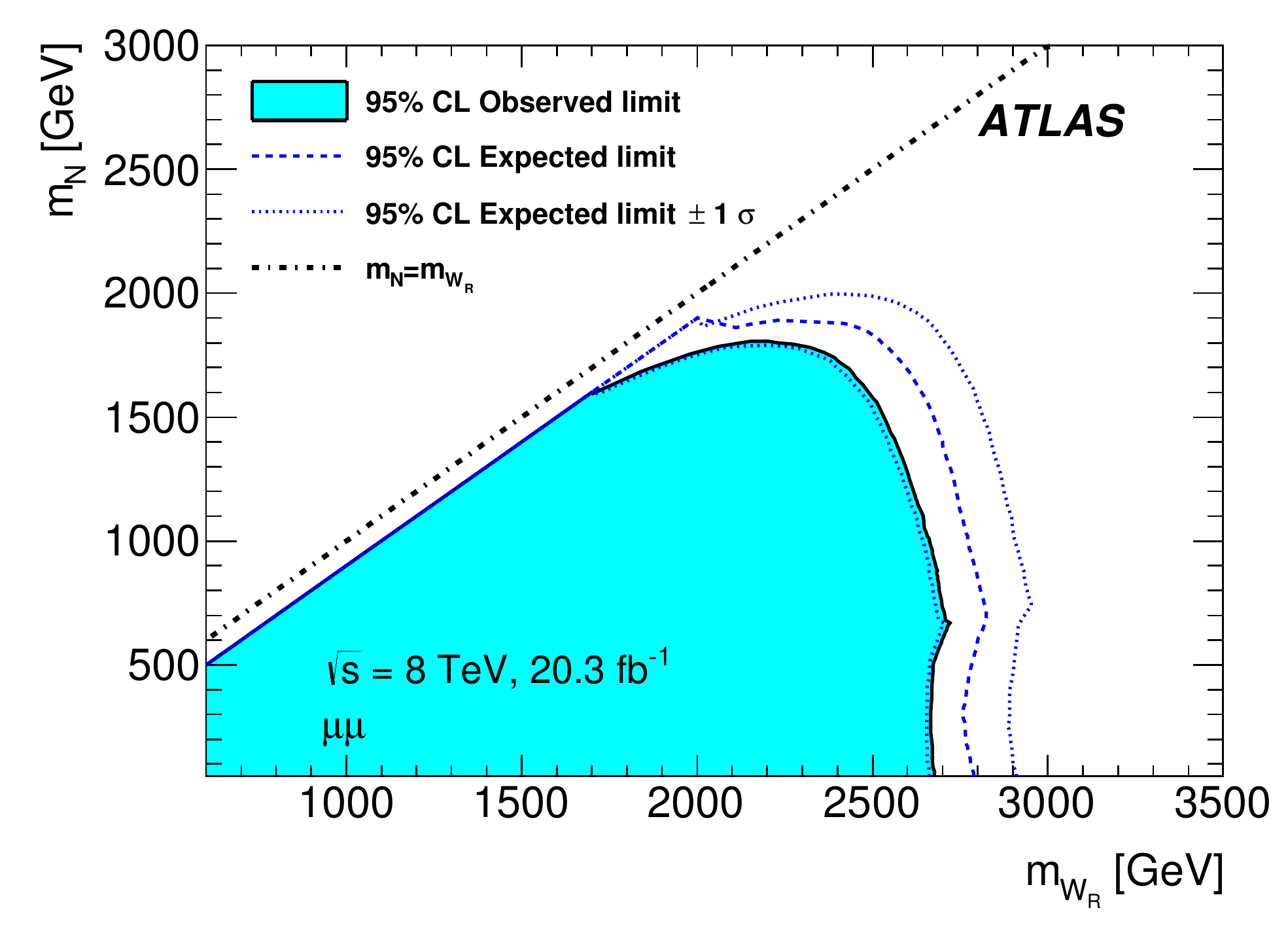}}
  \subfigure[]{\includegraphics[width=0.48\textwidth]{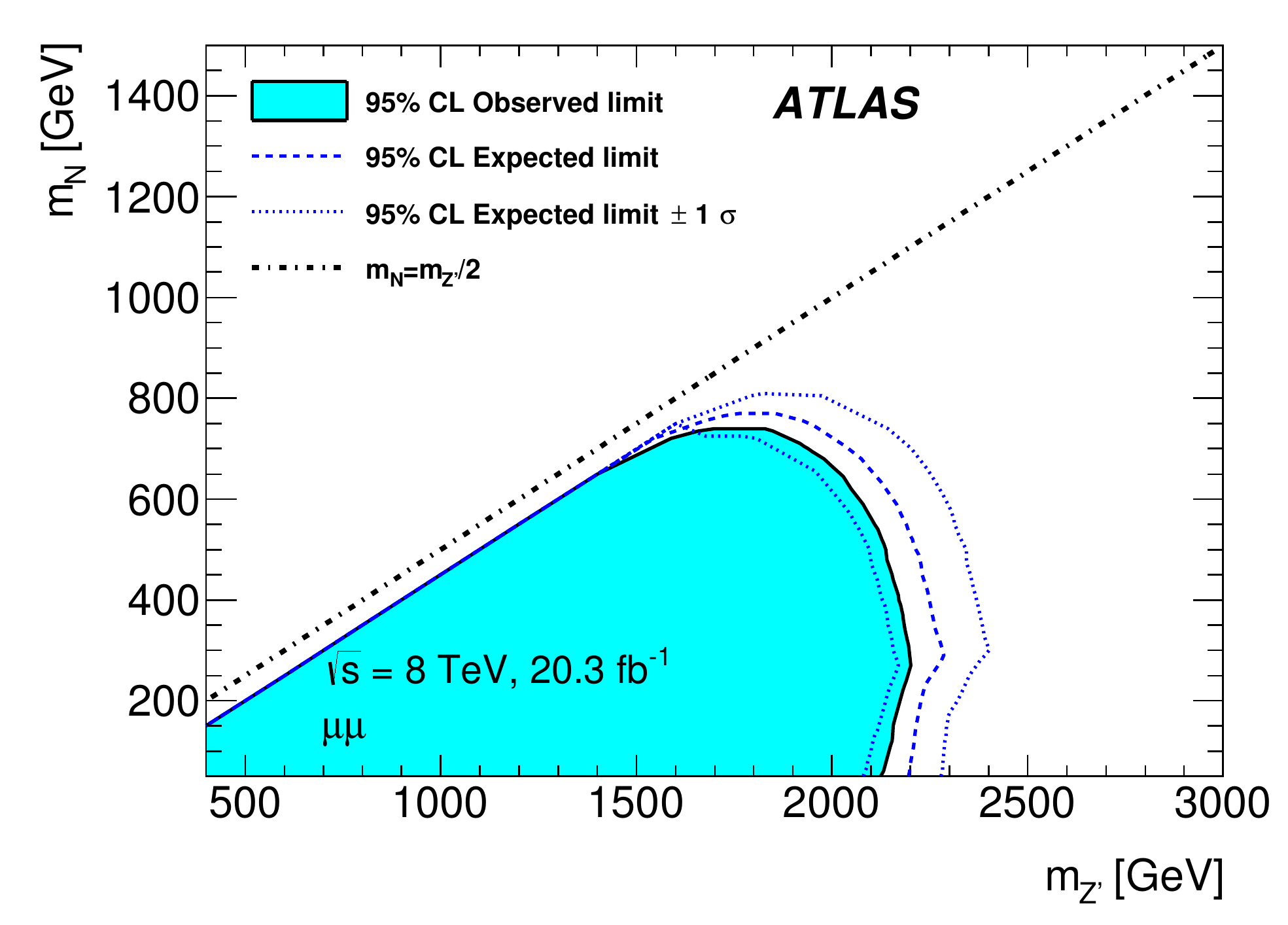}}
    \subfigure[]{\includegraphics[width=0.48\textwidth]{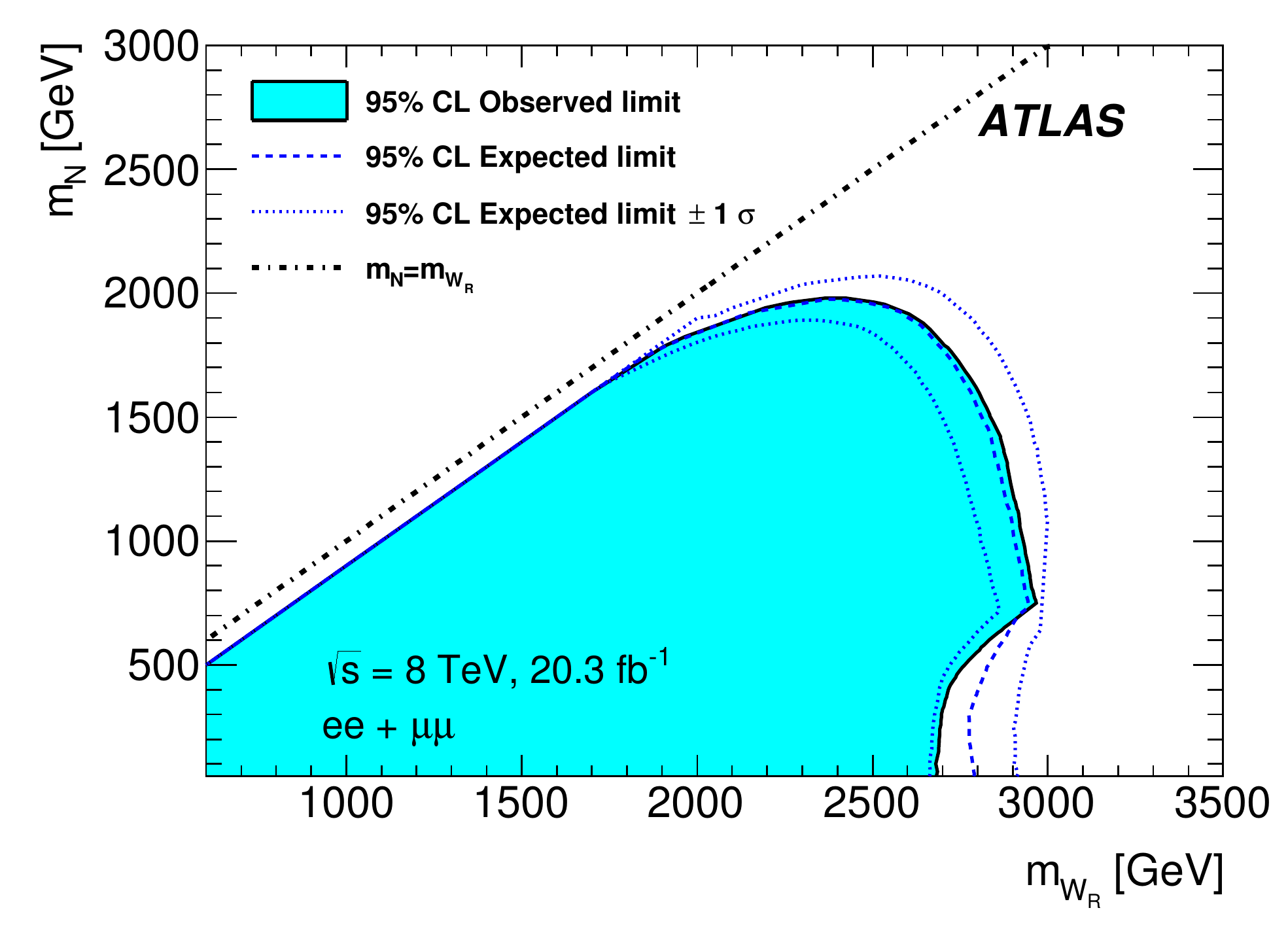}}
  \subfigure[]{\includegraphics[width=0.48\textwidth]{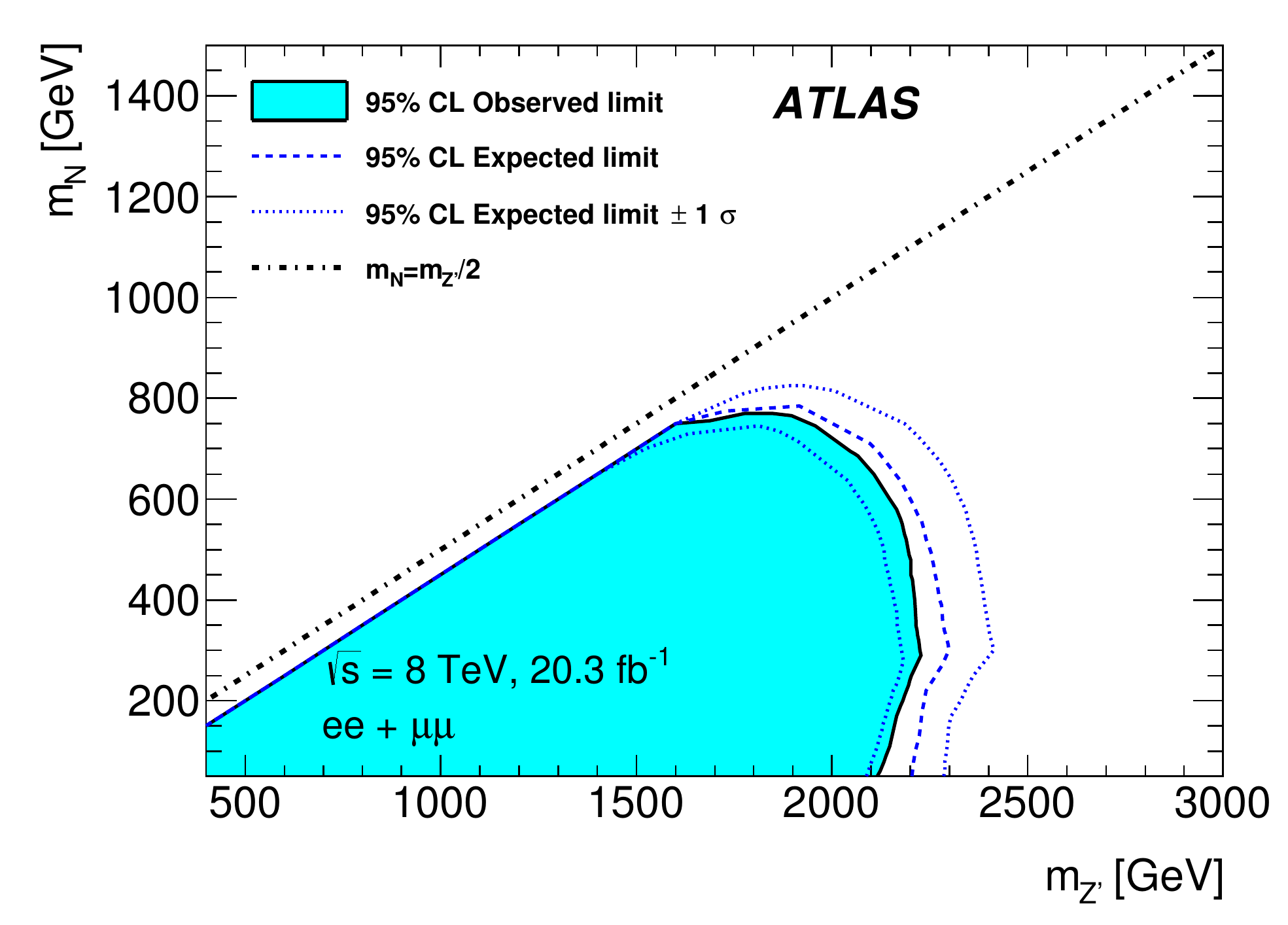}}
 \caption{Observed and expected exclusion contour at $95\%$~confidence level as a function of the mass of a heavy Majorana neutrino and of a $\WR$ (left) or $\ZR$~boson (right) within
  the LRSM. The limits in (a) and (b) show the scenario where the heavy neutrino has electron flavour and those in (c) and (d) show the scenario where it
  has muon flavour. The limits in (e) and (f) show the case of two degenerate neutrinos, one has electron flavour, and the other muon flavour (no mixing between lepton flavours is assumed).}
  \label{fig:LRSM_limit_Majorana_SS}
\end{figure}

\clearpage
\section{Conclusions}

The proton--proton collision data sample with a centre-of-mass energy of 8~TeV collected by ATLAS with an integrated luminosity of $20.3~\ifb$~has been used to search for the production of heavy Majorana neutrinos.
The selected events contain two same-sign charged leptons and high-$\pt$ jets. 
Two final selection criteria were used to provide sensitivity to two different benchmark models, the first being a minimal extension of the
SM, mTISM, and the second being a left-right symmetric extension of the SM, LRSM.
The background expectation in the signal regions is estimated using a combination of data-driven methods
and MC simulation.

No significant excess of events is observed above the SM background, and limits are set for each model.
 In the framework of a mTISM,
limits are set on the cross-section times branching ratio and
the mixing between heavy neutrinos and SM neutrinos as a function of the heavy neutrino mass. The observed limits exclude mixing parameters as low as
 $|V_{\mu N}|^2=0.0028$~and $|V_{e N}|^2=0.029$.
In the framework of an LRSM, limits are set
on heavy Majorana neutrino masses in the range 50$\GeV$ to 2000$\GeV$ for heavy gauge boson masses above 400$\GeV$. 
\\

\ackname

We thank CERN for the very successful operation of the LHC, as well as the
support staff from our institutions without whom ATLAS could not be
operated efficiently.

We acknowledge the support of ANPCyT, Argentina; YerPhI, Armenia; ARC,
Australia; BMWFW and FWF, Austria; ANAS, Azerbaijan; SSTC, Belarus; CNPq and FAPESP,
Brazil; NSERC, NRC and CFI, Canada; CERN; CONICYT, Chile; CAS, MOST and NSFC,
China; COLCIENCIAS, Colombia; MSMT CR, MPO CR and VSC CR, Czech Republic;
DNRF, DNSRC and Lundbeck Foundation, Denmark; EPLANET, ERC and NSRF, European Union;
IN2P3-CNRS, CEA-DSM/IRFU, France; GNSF, Georgia; BMBF, DFG, HGF, MPG and AvH
Foundation, Germany; GSRT and NSRF, Greece; RGC, Hong Kong SAR, China; ISF, MINERVA, GIF, I-CORE and Benoziyo Center, Israel; INFN, Italy; MEXT and JSPS, Japan; CNRST, Morocco; FOM and NWO, Netherlands; BRF and RCN, Norway; MNiSW and NCN, Poland; GRICES and FCT, Portugal; MNE/IFA, Romania; MES of Russia and NRC KI, Russian Federation; JINR; MSTD,
Serbia; MSSR, Slovakia; ARRS and MIZ\v{S}, Slovenia; DST/NRF, South Africa;
MINECO, Spain; SRC and Wallenberg Foundation, Sweden; SER, SNSF and Cantons of
Bern and Geneva, Switzerland; NSC, Taiwan; TAEK, Turkey; STFC, the Royal
Society and Leverhulme Trust, United Kingdom; DOE and NSF, United States of
America.

The crucial computing support from all WLCG partners is acknowledged
gratefully, in particular from CERN and the ATLAS Tier-1 facilities at
TRIUMF (Canada), NDGF (Denmark, Norway, Sweden), CC-IN2P3 (France),
KIT/GridKA (Germany), INFN-CNAF (Italy), NL-T1 (Netherlands), PIC (Spain),
ASGC (Taiwan), RAL (UK) and BNL (USA) and in the Tier-2 facilities
worldwide.

\bibliographystyle{atlasBibStyleWithTitle}
\bibliography{heavyN}
\newpage


\begin{flushleft}
{\Large The ATLAS Collaboration}

\bigskip

G.~Aad$^{\rm 85}$,
B.~Abbott$^{\rm 113}$,
J.~Abdallah$^{\rm 152}$,
S.~Abdel~Khalek$^{\rm 117}$,
O.~Abdinov$^{\rm 11}$,
R.~Aben$^{\rm 107}$,
B.~Abi$^{\rm 114}$,
M.~Abolins$^{\rm 90}$,
O.S.~AbouZeid$^{\rm 159}$,
H.~Abramowicz$^{\rm 154}$
% ...
\bigskip
\\
% ...
$^{11}$ Institute of Physics, Azerbaijan Academy of Sciences, Baku, Azerbaijan\\
% ...
$^{85}$ CPPM, Aix-Marseille Universit{\'e} and CNRS/IN2P3, Marseille, France\\
% ...
$^{90}$ Department of Physics and Astronomy, Michigan State University, East Lansing MI, United States of America\\
% ...
$^{107}$ Nikhef National Institute for Subatomic Physics and University of Amsterdam, Amsterdam, Netherlands\\
% ...
$^{113}$ Homer L. Dodge Department of Physics and Astronomy, University of Oklahoma, Norman OK, United States of America\\
$^{114}$ Department of Physics, Oklahoma State University, Stillwater OK, United States of America\\
% ...
$^{117}$ LAL, Universit{\'e} Paris-Sud and CNRS/IN2P3, Orsay, France\\
% ...
$^{152}$ Institute of Physics, Academia Sinica, Taipei, Taiwan\\
% ...
$^{154}$ Raymond and Beverly Sackler School of Physics and Astronomy, Tel Aviv University, Tel Aviv, Israel\\
% ...
$^{159}$ Department of Physics, University of Toronto, Toronto ON, Canada\\
% ...
\end{flushleft}


\end{document}